\documentclass{aa}

\usepackage{multirow}
\usepackage{comment}
\usepackage{graphicx}
\usepackage{circledsteps}
\usepackage{longtable}
\usepackage{txfonts}
\usepackage{lscape}

\usepackage[]{hyperref}

\begin{document} 

\title{Internal rotation and buoyancy travel time of 60 $\gamma$~Doradus stars from uninterrupted TESS light curves spanning 352\,days\thanks{The extended Tables in the Appendix and the analysed data sets are also available in electronic form at \href{https://doi.org/10.5281/zenodo.7218038}{https://doi.org/10.5281/zenodo.7218038}.}}

\subtitle{}

\author{
    S.~Garcia\inst{\ref{ivs}}
    \and
    T.~Van Reeth\inst{\ref{ivs}}        
    \and
    J.~De Ridder\inst{\ref{ivs}}
    \and
    C.~Aerts\inst{\ref{ivs},2,3}
}

\institute{
    Institute of Astronomy, KU Leuven, Celestijnenlaan 200D, B-3001 Leuven, Belgium \\
    \email{timothy.vanreeth@kuleuven.be}
    \label{ivs}
    \and
Department of Astrophysics, IMAPP, Radboud University Nijmegen, PO Box 9010,
6500 GL, Nijmegen, The Netherlands
\and 
Max Planck Institute for Astronomy, Koenigstuhl 17, 69117, Heidelberg, Germany
    }

\date{Received month dd, yyyy; accepted month dd, yyyy}

% \abstract{}{}{}{}{} 
% 5 {} token are mandatory
 
  \abstract
   {Gamma Doradus (hereafter $\gamma$~Dor) stars are gravity-mode pulsators whose periods carry information about the internal structure of the star. These periods are especially sensitive to the internal rotation and chemical mixing, two processes that are currently not well constrained in the theory of stellar evolution.}
   {We aim to identify the pulsation modes and deduce the internal rotation and buoyancy travel time for 106 $\gamma$~Dor stars observed by the Transiting Exoplanet Survey Satellite (TESS) mission in its southern continuous viewing zone (hereafter S-CVZ). We rely on 140 previously detected period-spacing patterns, that is, series of (near-)consecutive pulsation mode periods.}
   {We used the asymptotic expression to compute gravity-mode frequencies for ranges of the rotation rate and buoyancy travel time that cover the physical range in $\gamma$~Dor stars. Those frequencies were fitted to the observed period-spacing patterns by minimising a custom cost function. The effects of rotation were evaluated using the traditional approximation of rotation, using the stellar pulsation code GYRE.}
   {We obtained the pulsation mode identification, internal rotation and buoyancy travel time for 60 TESS $\gamma$~Dor stars. For the remaining 46 targets, the detected patterns are either too short or contained too many missing modes for unambiguous mode identification, and longer light curves are required. For the successfully analysed stars, we found that period-spacing patterns from 1-yr long TESS light curves can constrain the internal rotation and buoyancy travel time to a precision of $\rm 0.03~d^{-1}$ and 400~s, respectively, which is about half as precise as literature results based on 4-yr {\it Kepler\/} light curves of $\gamma$~Dor stars.}
{}


\titlerunning{Internal rotation and buoyancy travel time of rotating dwarfs}
\authorrunning{Garcia et al.}

\maketitle
%

%--------------------------------------
\section{Introduction}
%--------------------------------------

After the {\it Kepler\/} satellite allowed us to  probe the interior of thousands of stars \citep{Koch2010}, it became clear that the theory of stellar evolution \citep[e.g.,][]{Kippenhahn2012} still needs extra tuning to explain the observed internal stellar structure. Stellar astrophysics has since then experienced a fast proliferation of competing internal stellar processes that aim to explain the differences between theory and observation, notably in terms of angular momentum transport and chemical mixing \citep[e.g.,][]{Rogers2015, Moravveji2015, Moravveji2016, Fuller2017, Townsend2018, Eggenberger2019, Ouazzani2019, Mombarg2022, Pedersen2021}. Those processes are currently being evaluated to identify the dominant ones in different stellar classes and at different stages of the stellar evolution. A critical factor in this research is the size of the samples of asteroseismically modelled pulsators in various evolutionary phases \citep[e.g.,][for extensive reviews]{Aerts2019,GarciaBallot2019,Corsico2019} and the amount of useful information per star. The available samples of dwarfs are currently too small to evaluate their complex and interacting processes of internal rotation and mixing, particularly for intermediate- and high-mass stars, with $M \gtrsim 1.4M_\odot$ \citep{Bowman2020c,Aerts2021}.

The ongoing Transiting Exoplanet Survey Satellite (TESS) mission has been providing the community with full-frame images, giving rise to hundreds of thousands of light curves of bright stars \citep{Ricker2015}. These TESS data may help to tackle the too small sample sizes of pulsating dwarfs with good asteroseismic modelling, which requires spectroscopy to get good fundamental parameters, while this is hard to establish for {\it Kepler\/} stars. As TESS continues to observe stars, the precision of measured pulsation mode frequencies increases. This will allow us to asteroseismically test future improvements of stellar structure and evolution theory from stellar pulsation computations. New methodological frameworks coupled to high-precision uninterrupted space photometric data of intermediate-mass gravity-mode pulsators have been developed, including those by \citet{Kurtz2014,Saio2015,Moravveji2015, Moravveji2016,VanReeth2016, Ouazzani2017,VanReeth2018, Aerts2018,Szewczuk2018,Ouazzani2019,LiGang2020, Pedersen2021,Michielsen2021,Bowman2021c,Saio2021,Szewczuk2022}. The powerful tools in these studies have been used to identify the stars with maximum asteroseismic scientific potential among the ever-increasing data provided by space missions.

The traditional approximation of rotation \citep[TAR;][]{Eckart1960,Townsend2003,Mathis2009} is an essential tool to perform gravity-mode asteroseismology of intermediate-mass stars, because they tend to be fast rotators, demanding the inclusion of the Coriolis acceleration into the treatment of the pulsation predictions. Ever more complex versions of the TAR, including the centrifugal and Lorentz forces, have been developed to allow for accurate predictions of gravito-inertial pulsation frequencies comparable with observations \citep{Prat2019a,Prat2020a,JordanVanBeeck2020,Henneco2021,Dhouib2021a,Dhouib2021b,Dhouib2022}. These various versions of the TAR inform us how gravity-mode pulsation frequencies of dwarfs are affected by rotation and magnetism and allow us to search for such signatures in the space photometry \citep{VanReeth2015a, Christophe2018, Szewczuk2021, Pedersen2021, Garcia2022}. In terms of TESS data, \citet{Garcia2022} relied on theoretical predictions to create a catalogue of 106 gravito-inertial pulsators with detected period-pacing patterns affected by rotation. This follow-up paper extracts the asteroseismic information from that catalogue with the aim to constrain the interior rotation and buoyancy travel time in those stars, based on $\sim$1\,year of uninterrupted space photometry. Our methodology is validated by analyzing a well-studied sample of {\em Kepler} $\gamma$~Dor stars, and comparing our results with those in the literature. 

Period-spacing patterns are series of periods of (near-)consecutive g~modes with identical surface geometry. The spacings between periods of modes with consecutive radial order  are important observational diagnostics. They offer to
assess information contained in the pulsation mode periods and to couple this information to internal physical processes. The slope of the pattern is a reflection of the internal rotation \citep{Bouabid2013,VanReeth2016,Ouazzani2017}, with the pulsators revealing steeper patterns being the faster rotators. In addition, the structure in the patterns (by means of dips or wiggles) indicates the presence of trapped modes produced by strong internal gradients influenced by microscopic and/or macroscopic internal mixing \citep{Miglio2008,Pedersen2018,Michielsen2019,Michielsen2021,Pedersen2021, Mombarg2022}. These processes are poorly understood \citep{SalarisCassisi2017}, but theoretical approximations of them based on the TAR and coupled to observed, identified gravito-inertial modes allow us to deduce the mixing and abundance profiles in the stellar interior, for a variety of global and local stellar parameters. Therefore we concentrate our efforts on finding gravity- and gravito-inertial modes (hereafter, collectively referred to as g modes) in pulsators with period-spacing patterns from TESS data. In this way, we compose a TESS sample with the potential to derive the profiles of internal rotation and mixing inside $\gamma$~Dor stars along the entire core-hydrogen burning phase.

Once the data are assembled, asteroseismic modelling is only as good as the  identification of the pulsation modes involved in the period-spacing patterns \citep[see][]{Bowman2021c}. Proper mode identification in fast rotators is a challenging problem. In this paper, we use a comprehensive approach to test the likelihood of possible low-degree mode identifications in the detected period-spacing patterns found by \citet{Garcia2022}. The plan of the paper is as follows: Section\,\ref{Sec:method} describes our methodology to determine the internal rotation and buoyancy travel time from period-spacing patterns. Section\,\ref{Sec:results} presents our asteroseismic catalogue of $\gamma$~Dor stars, discusses some of their basic properties, and identifies the stars with maximum asteroseismic scientific potential. Section\,\ref{Sec:spec_characterization} characterises our catalogue with spectroscopic quantities. Section\,\ref{Sec:Kepler_stars} validates our methodology by a re-analysis of a sample of {\it Kepler\/}. Finally, Section\,\ref{Sec:discussion} presents our main conclusion.  

%--------------------------------------
\section{Interior rotation and buoyancy travel time from g-mode period-spacing patterns}\label{Sec:method}
%--------------------------------------

For a non-rotating non-magnetic star, the surface geometry of a stellar pulsation mode is characterised by a spherical harmonic function with two quantum numbers: the angular degree $\ell$ and the azimuthal order $m$. Aside from $\ell$ and $m$, the radial order $n$ correlates with the number of radial nodes of the displacement vector of the mode \citep[e.g.,][]{Aerts2010}.
For such a simplified approximation of a star, pulsation periods of g modes in the asymptotic regime, that is when the radial order $n\gg \ell$, depend on the buoyancy travel time across the mode cavity, as follows:

\begin{equation} \label{Eq:buoyancy_radius}
    \Pi_0 = \frac{2\pi^2}{\int_{r_1}^{r_2}N(r)\,r^{-1}\,\mathrm{d}r}\, ,
\end{equation}
with $N$ being the Brunt–V\"ais\"al\"a frequency defined in a cavity with inner and outer boundaries $r_1$ and $r_2$, respectively. 
Note that Eq. (\ref{Eq:buoyancy_radius}) implicitly involves many details of the internal structure of the star, notably the local density $\rho$, gravity $g$, pressure $p$, temperature gradients $\nabla\equiv ({\rm d}\ln T/{\rm d}\ln p)$ and its adiabatic version $\nabla_{\rm ad}$, as well as the gradient of the mean molecular weight $\nabla_\mu$ \citep[e.g.,][]{Aerts2010}:
\begin{equation} \label{Eq:BV}
    N(r)\simeq \frac{g^2\rho}{p} \left(\nabla_{\rm ad}-\nabla+\nabla_\mu\right)\,.
\end{equation}
In this regime of stellar oscillations in a non-rotating non-magnetic star, its radial-order series of g-mode periods are equally spaced as
\begin{equation} \label{Eq:asymptomatic_approx}
P_{n \ell m} = \frac{\Pi_0}{\sqrt{\ell(\ell+1)}}\, (n+\alpha_g),
\end{equation}
where $\alpha_g$ is a phase term. 

If the star is rotating, the pulsation frequencies are affected by the Coriolis and centrifugal forces. In the current work, we neglect the centrifugal force and assume rigid rotation with frequency $f_{\rm rot}$. The effect of the Coriolis force on the g modes is described well by the traditional approximation of rotation \citep[TAR;][]{Eckart1960,Townsend2003,Mathis2009}. The geometry of stellar pulsations in rotating stars is  characterised by the Hough functions, which are solutions to the Laplace tidal equations. Thus, the oscillation modes can no longer be described by spherical harmonics with $\ell$ and $m$. Rather, the eigenvalues associated with the Hough functions are $\lambda_{\ell m s}$, which depend on the spin parameter defined as
\begin{equation}
    s \equiv \frac{2f_{\textrm{rot}}}{f_{n \ell m}^{\textrm{co}}},
\end{equation}
with $f_{n \ell m}^{\rm co}$ the mode frequency in the co-rotating reference frame with rotation frequency $f_{\rm rot}$. The TAR then predicts a radial-order series of g-mode periods as
\begin{equation} \label{Eq:TAR}
    P_{n \ell m}^{\textrm{co}} = \frac{\Pi_0}{\sqrt{\lambda_{\ell m s}}}(n+\alpha_g).
\end{equation}
In the inertial reference frame of an observer, this series is measured as \citep[e.g.,][]{VanReeth2016}
\begin{equation}\label{Eq:inertial_frame}
    P_{n \ell m}^\textrm{in} = \frac{1}{1/P_{n \ell m}^{\textrm{co}} + mf_\textrm{rot}}.
\end{equation}

Where modes with $m > 0$ and $m < 0$ are prograde and retrograde modes, respectively.

Corrections to theoretical predictions of g-mode periods due to magnetic fields, the centrifugal force, and differential rotation have been explored recently in \citet{JordanVanBeeck2020}, \citet{Dhouib2022}, \citet{Henneco2021}, and \citet{Dhouib2021b}, relying on the theoretical frameworks developed by  \citet{Prat2019a,Prat2020a} and \citet{Dhouib2021a}, respectively. As with the {\it Kepler\/} data exploitation of g-mode pulsators so far, this paper will neglect these improvements of the TAR for a first exploration. Indeed, our aim is to assess the opportunities of TESS for the exploitation of g-mode patterns. Moreover, while the additional pulsation frequency shifts caused by these neglected effects are formally detectable within the available space-based photometric observations, they are two orders of magnitude smaller than the pulsation frequency shifts caused by the Coriolis acceleration \citep{Henneco2021,Dhouib2021b}. Thus, we can safely limit this work to interpretations based on the simplified version of the TAR via Eq.\,(\ref{Eq:TAR}). The neglected processes are not expected to significantly affect the obtained mode identifications or rotation rate measurements. This also allows for faster calculations for the many period-spacing patterns in our catalogue compared to the more complex dispersion relations in the above-mentioned theory papers.

\subsection{Methodological framework for the estimation of the interior rotation and buoyancy travel time}

The input catalogue of this study contains 106 $\gamma$~Dor stars observed by  the TESS space telescope in its Cycle 1 in the Southern ecliptic hemisphere. These 106 g-mode pulsators exhibit 140 period-spacing patterns as reported by \citet{Garcia2022}. The patterns were detected in the 1-yr light curves of the S-CVZ. Unlike the 4-yr light curves assembled by the {\em Kepler\/} space telescope, where missing modes are infrequent thanks to the long time base \citep[e.g.,][]{VanReeth2015b}, the 1-yr light curves used in this work are more likely to result in sparse observed pulsation patterns, that is, some g-mode periods in the radial-order series given by Eq. (\ref{Eq:inertial_frame}) may not be detected.

In this work, we fitted the detected g-mode periods, $P_{n \ell m}^{\rm \;obs}$, directly. We calculated theoretical predictions for the g-mode periods, $P_{n \ell m}^\textrm{in}$ (hereafter simplified as $P_{n \ell m}$), in Eq.\,(\ref{Eq:inertial_frame}). The effects of rotation are evaluated using the TAR module of GYRE\footnote{\href{http://gyre.readthedocs.io}{http://gyre.readthedocs.io}} \citep[v6,][]{Townsend2013,Townsend2018,Goldstein2020}. It provides tabulated numerical solutions to the Laplace tidal equation, which were obtained with the spectral matrix method, as described by \citet{Townsend2003b}. In Section\,\ref{Sec:Kepler_stars}, we validate our method by evaluating $\gamma$~Dor stars that have been studied in the literature before, and comparing the results.

The comparison between g-mode periods from asymptotic theory and the observations was done by adapting the cost function from \citet[][]{Garcia2022} to the parameters to be estimated as
\begin{equation}  \label{Eq:cost_function}
S \left(\Pi_0, f_{\rm rot}, \alpha_g \right)=\sum_{n} \; \frac{A_{n}^{\rm obs}}{A_{\rm max}^{\rm obs}} \; \frac{\left(P_{\,n}^{\rm\;obs}-{P}_{n}\right)^{2}}{\sigma_{P_n^{\rm \;obs}}^{2} + \Delta {P}_{n}^{\;2}}\;,
\end{equation}
where $P_n$ is the theoretical mode period and $\Delta P_n$ its local period spacing, $P^{\rm \;obs}_n$ is the observed mode period, $\sigma_{P_n^{\rm \;obs}}$ its uncertainty, $A_n^{\rm obs}$ its observed amplitude, and $A_{\rm max}^{\rm obs}$ is the maximum detected amplitude in the observed Lomb-Scargle periodogram as computed by \citet[][]{Garcia2022}. The angular wavenumbers, $\ell$ and $m$, are omitted in Eq.\,(\ref{Eq:cost_function}) for readability. Identification of the radial orders of the observed mode periods, $P_{\,n}^{\rm\;obs}$, with respect to the theoretically predicted period sequence, $P_{\,n}$, was done by finding the closest match between the dominant frequency in the observed sequence and the predicted one. In doing so we included the information about missing radial orders in the observed sequences as reported in \citet[][]{Garcia2022} by setting their corresponding amplitude to zero. This step is paramount. Because the asymptotic model for the period-spacing pattern modelling has three free parameters, an observed pattern has to contain at least four individual pulsation modes for a successful analysis. If there are too many missing radial orders or the number of missing radial orders is uncertain, the minimum number of observed modes for a successful analysis increases.

\subsection{Identification of the angular wavenumbers}\label{Sec:mode_identification}

The identification of the angular wavenumbers deduced from the observed g-mode patterns was done following the labelling system introduced by \citet{Lee1997}, where gravito-inertial and Rossby modes of consecutive radial order $n$ are labelled by the set of indices $(k,m)$, with $k\equiv \ell-|m| \ge 0$ for gravito-inertial modes and $k<0$ for Rossby modes. We computed the mode periods from the TAR for all dipole and quadrupole modes, as well as the Rossby modes $(-2,-1)$ and $(-1,-1)$. This results in ten different g-mode patterns $(k,m)$, each of them returning a minimum value of the cost function denoted by $S_{\rm min}^{km}$. 

To deduce the most likely identification for each detected g-mode pattern, we considered the coarse grid with parameter ranges for $\Pi_0$, $f_{\rm rot}$, and $\alpha_{\rm g}$ as indicated in Table\,\ref{Tab:parameter_space}. These ranges for $\Pi_0$ and $f_{\rm rot}$ are wider compared to the ones reported in the literature for the large sample of {\it Kepler\/} $\gamma$~Dor pulsators \citep[e.g.,][]{VanReeth2015b,LiGang2020}. We thus expect all best-fit values to show well-defined minima in the cost function. The phase term $\alpha_g$ in Eq. (\ref{Eq:asymptomatic_approx}) and (\ref{Eq:TAR}) captures the effect of boundary layers inside the star and may attain any value \citep{Aerts2010}. However, we limit its values in Eq. (\ref{Eq:cost_function}) to the effective range $0 \le \alpha_g \le 1$. In other words, we would interpret a value of $\alpha_g>1$ and $\alpha_g<0$ as belonging to a mode with the overtone $n+1$ and $n-1$, respectively. When the cost function changes monotonically within the parametric range, leading to a solution at the edges of the range, we reject that solution for the mode identification. 

\begin{figure}
\centering
\includegraphics[width=\hsize,clip]{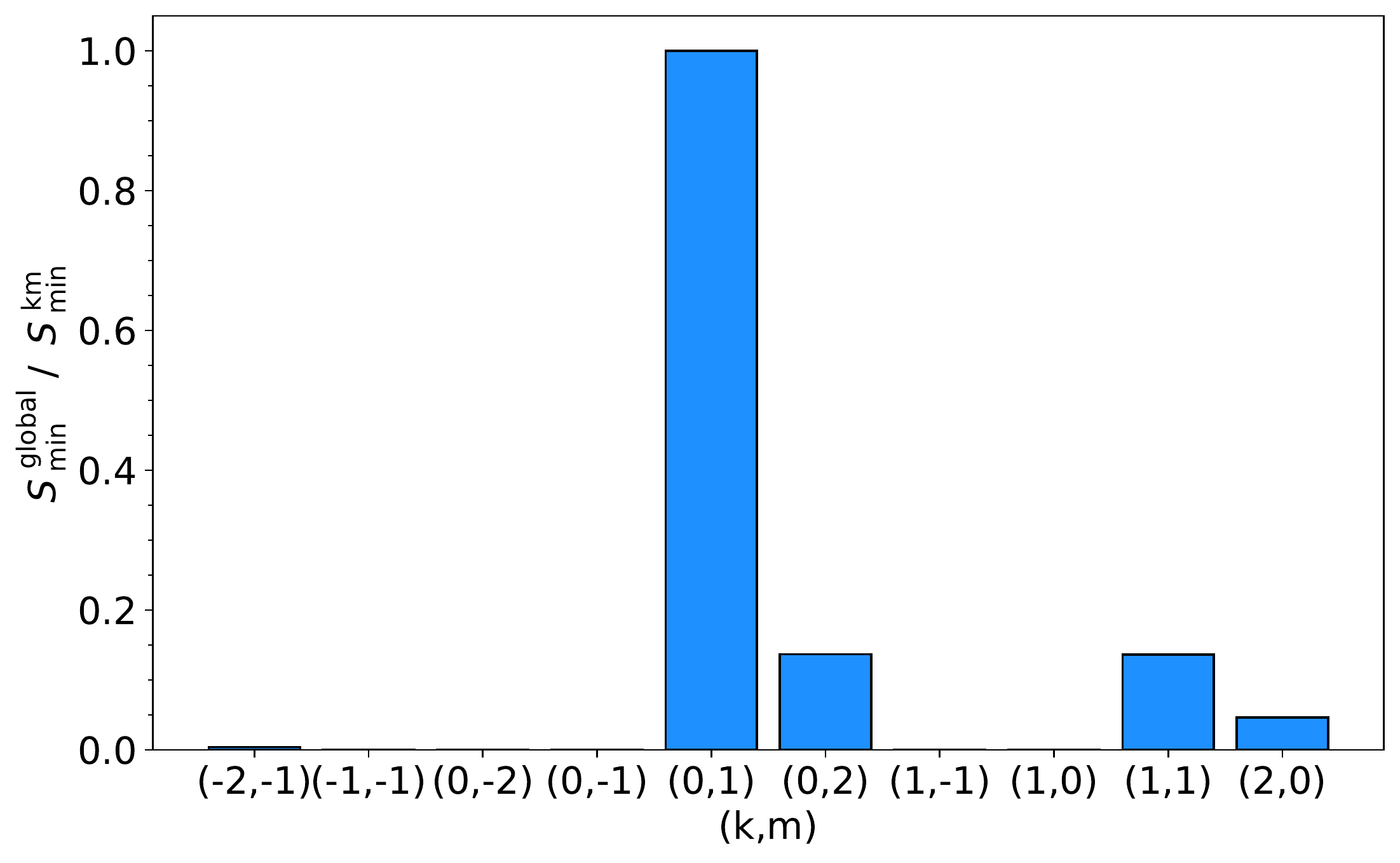}
\caption{Mode identification deduced from the main period-spacing pattern detected for TIC\,381950897.}
\label{Fig:mode_id}
\end{figure}

\begin{figure}
\centering
\includegraphics[width=\hsize,clip]{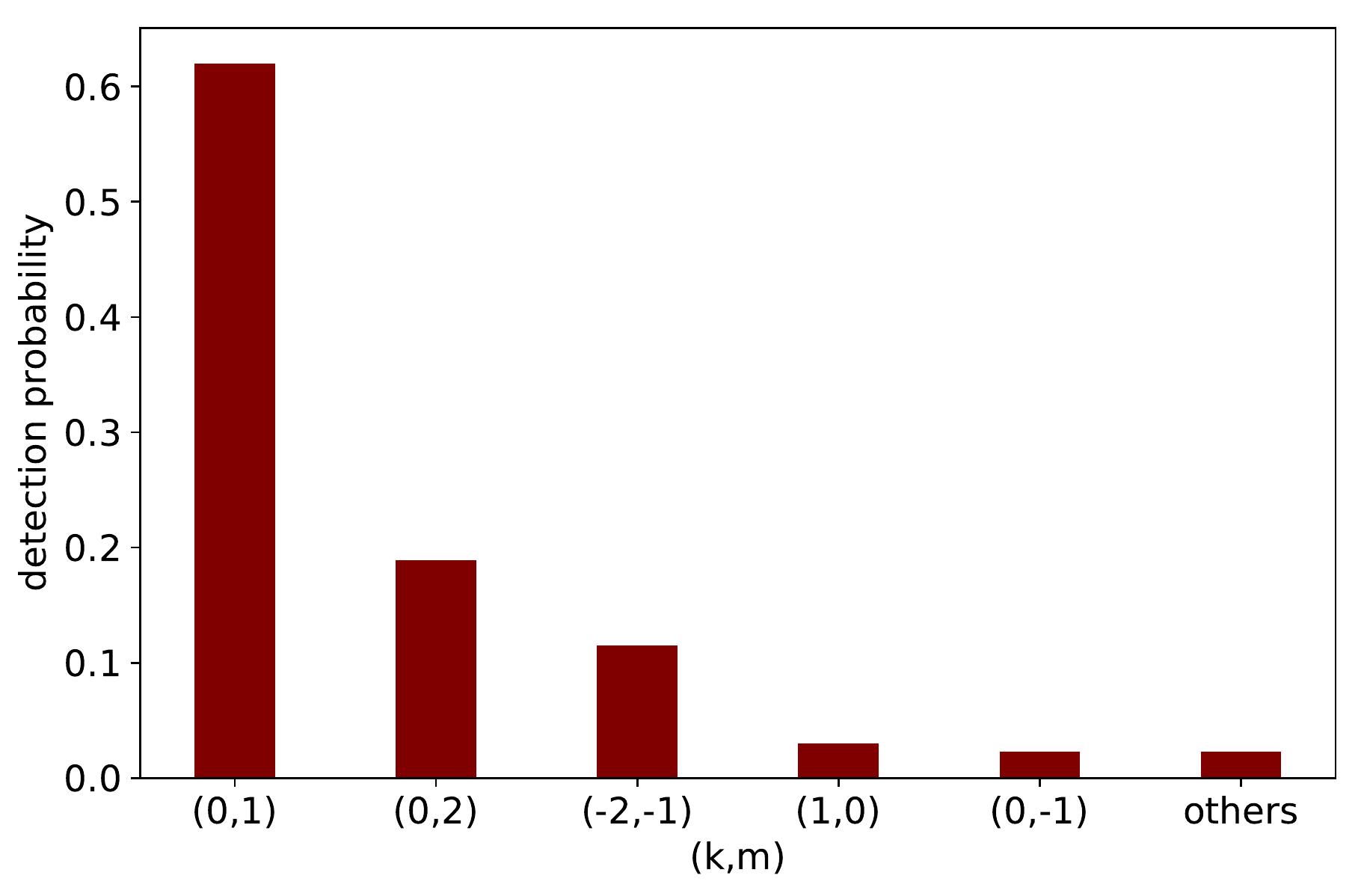}
\caption{Probability distribution for the detection of different g-mode pulsation geometries $(k,m)$, based on the observed sample distribution from the $\gamma$\,Dor pulsator study by \citet{LiGang2020}.}
\label{fig:Li2020_probabilities}
\end{figure}

We adopted the global minimum among all the combinations, $S_{\rm min}^{\rm global}$, as the most likely mode identification. This identification process is exemplified in Fig.\,\ref{Fig:mode_id} by using the main period-spacing pattern present in TIC\,381950897, that is, the one with the dominant frequency. This period-spacing pattern has eight detected g-mode periods and eight missing mode periods, which is a common case in the catalogue. The figure shows the ratio $S_{\rm min}^{\rm global}/S_{\rm min}^{km}$ for each mode combination. The relative heights of the bars represent how likely the identification of $(k,m)$ is for the detected mode pattern of this star among the considered options. Whenever the identification was ambiguous, that is if multiple bars in the identification diagnostic tool as the one shown in Fig.\,\ref{Fig:mode_id} had a height close to unity, we manually selected the most likely identification, based on the sample distribution of the different mode geometries that were observed and reported by \citet{LiGang2020}. These authors detected and identified 960 period-spacing patterns in a sample of more than 600 $\gamma$~Dor pulsators from 4-yr {\it Kepler\/} light curves, and summarised the relative sample fractions of the different mode identifications in their Fig.\,6 (represented in Fig.\,\ref{fig:Li2020_probabilities} in this work). Where necessary, we used this distribution to assess the probability of observing different mode geometries. For example, as already shown by \citet{LiGang2020}, we expect sectoral modes ($k=0$) to be more likely to be detected because rapid rotation tends to reduce the photometric visibility of the oscillations by confining the pulsational variability to the equatorial regions \citep{Townsend2000,Dhouib2022}. Because this effect is less strong for prograde sectoral modes, they have a higher visibility in most stars.

\begin{figure}
\centering
\includegraphics[width=\hsize,clip]{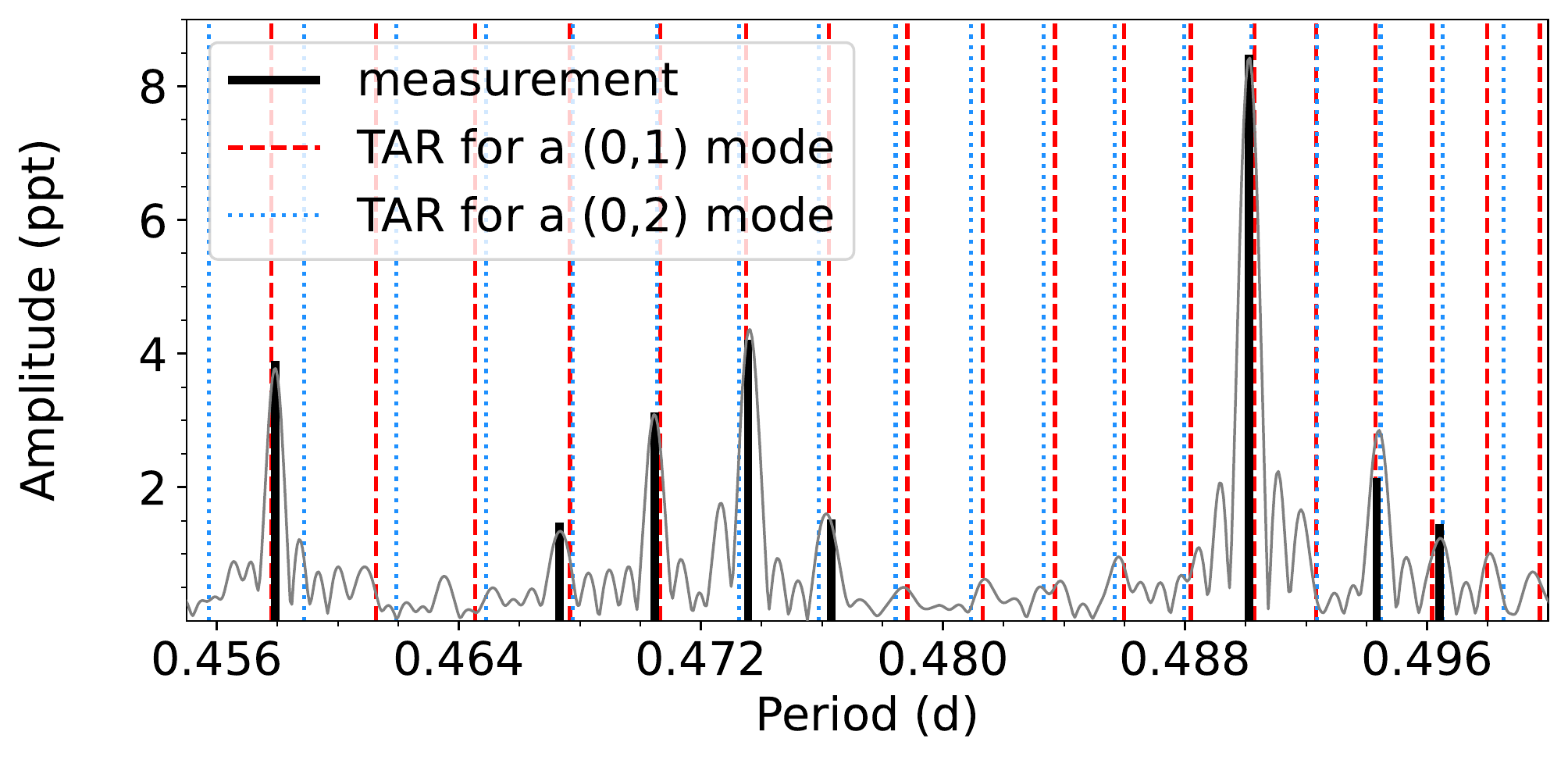}
\caption{TAR solution for the first and second most plausible pulsation mode identifications, that is $(0,1)$ and $(0,2)$, resulting from the main detected g-mode pattern of TIC\,381950897 from the distribution in Fig.\,\ref{Fig:mode_id}. The grey curve is the Lomb-Scargle periodogram.}
\label{Fig:two_mode_ids}
\end{figure}

Figure\,\ref{Fig:two_mode_ids} shows the TAR best-fit solution for the period-spacing patterns generated with the two most likely pulsation mode identifications in Fig.\,\ref{Fig:mode_id}, namely $(k,m)=(0,1)$ and $(0,2)$. The two solutions converged to reproduce the mode with the highest amplitude, occurring with a period of about 0.49\,d, optimally. This is a consequence of using the detected mode amplitudes as weights in the cost function defined in Eq.\,(\ref{Eq:cost_function}).

\begin{figure*}
\resizebox{\hsize}{!}
{\includegraphics[width=\hsize,clip]{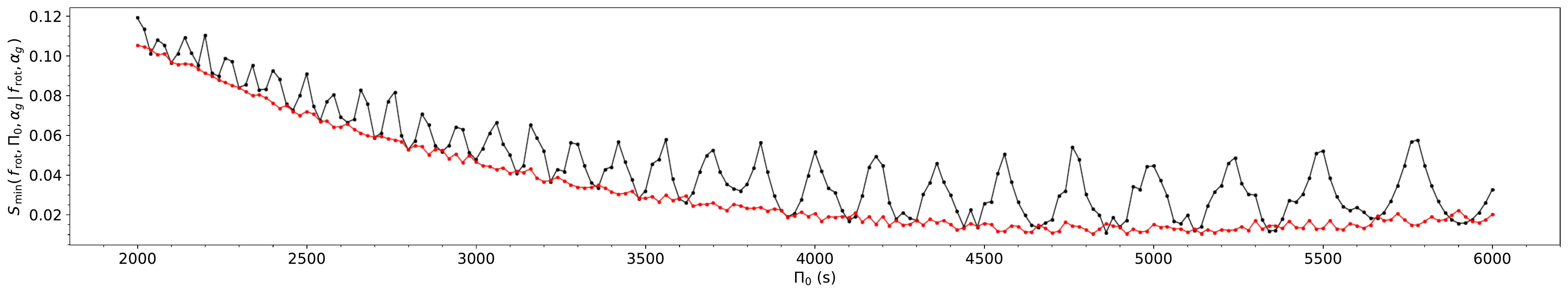}}
\caption{Effect of the resolution of $f_{\rm rot}$ in the parameter space on the determination of $\Pi_0$ for TIC\,381950897. The black  and red curves represent the cost function evaluated on the sparse and dense parameter space indicated in Table\,\ref{Tab:parameter_space}, respectively.}
\label{Fig:grids}
\end{figure*}

\begin{table}
\centering
\caption{3D space considered for the mode identification and parameter estimation.}
\begin{tabular}{lrrr}
 \hline \hline
                       & $\Pi_0$  & $f_{\rm rot}$  & $\alpha_{\rm g}$ \\
                       & (s) &  (d$^{-1}$) &  \\ \hline
min value        & 2000    & 0.000     &  0.000 \\
max value        & 6000    & 3.000     &  1.000\\
small step             & 20      & 0.002  & 0.125  \\
large step             & 20      & 0.010  & 0.125  \\ \hline
\end{tabular}
\tablefoot{The large step was used to generate a coarse parameter space for the mode identification while the small step was subsequently used to generate a dense parameter space to refine the estimation of $f_{\rm rot}$ and $\Pi_0$.}
\label{Tab:parameter_space}
\end{table}

\begin{figure*}
\resizebox{\hsize}{!}
{\includegraphics[width=\hsize,clip]{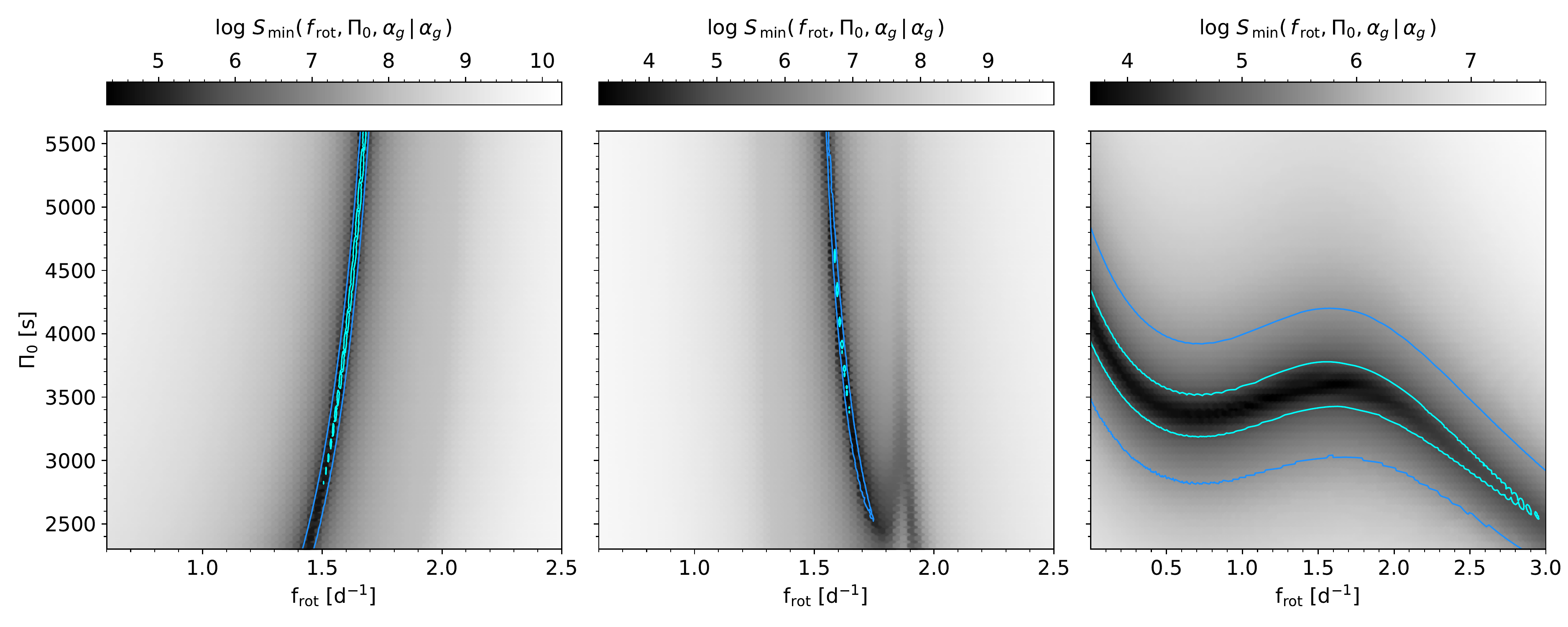}}
\caption{Cost function evaluated for the parameter
ranges for the internal rotation and buoyancy travel time. Each panel shows a typical correlation between $f_{\rm rot}$ and $\Pi_0$. Left: main pattern in TIC\,381950897 with mode identifications (0,1). Middle: secondary pattern in TIC\,381950897 with mode identifications (-2,-1). Right: pattern of TIC\,293221812 with mode identifications (0,-1). The blue contour lines indicates $S=100\,S_{\rm min}$ while the cyan one indicates $S=10\,S_{\rm min}$.}
\label{Fig:Pi0_vs_frot}
\end{figure*}

\subsection{Estimation of $f_{\rm rot}$ and $\Pi_0$}

Once the pulsation pattern was identified, we refined the estimates of $\Pi_0$ and $f_\textrm{rot}$ by searching for a minimum of the cost function $S$ in a denser parameter space also indicated in Table\,\ref{Tab:parameter_space}. For computational reasons, this denser space was only considered for the final mode identification. The impact of the resolution of $f_{\rm rot}$ on the estimation of $\Pi_0$ is shown in Fig.\,\ref{Fig:grids} for the main pattern in TIC\,381950897. The many local minima in black are due to the finite resolution of the grid and the strong correlation between $\Pi_0$ and $f_\textrm{rot}$. This is illustrated in Fig.\,\ref{Fig:Pi0_vs_frot}. The left panel represents a prograde mode, while the middle panel represents a retrograde mode. These panels indicate that a change in $\Pi_0$ can be compensated by a small change in $f_\textrm{rot}$. Proper estimation of both parameters therefore requires a sufficiently dense parameter grid, particularly for $f_{\rm rot}$,  and proper treatment of degeneracies in the computation of the uncertainties. The black curve in Fig.\,\ref{Fig:grids} results from the rough sampling for $f_\textrm{rot}$, while the red curve stems a five times denser grid step. As a result of the denser 3D grid search, the estimation of $\Pi_0$ results from a smoother curve without numerous local minima. 

\begin{figure}
\centering
\includegraphics[width=\hsize,clip]{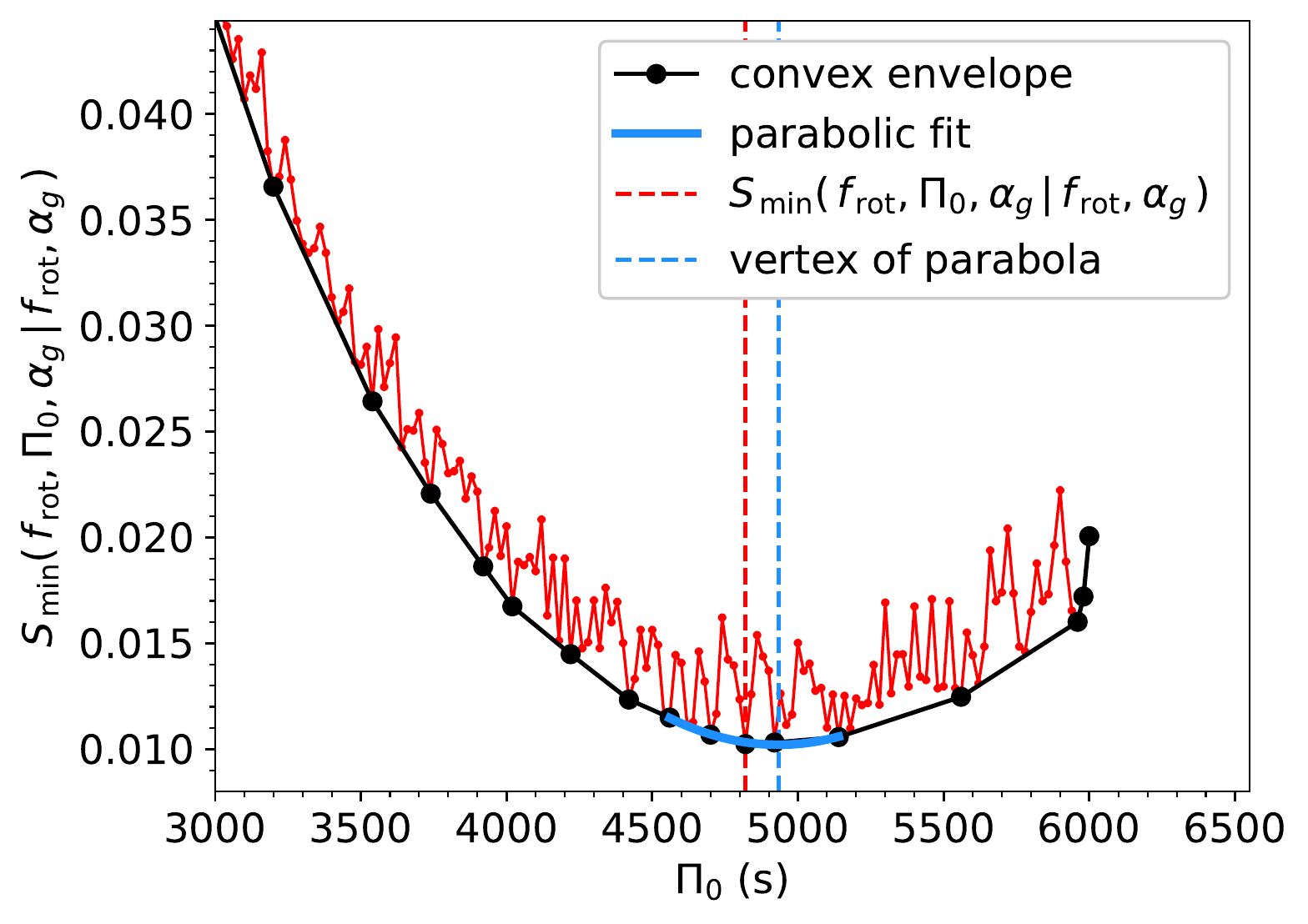}
\caption{Further refinement of the solution shown in Fig.\,\ref{Fig:grids} for TIC\,381950897. The red curve is the same as in Fig.\,\ref{Fig:grids}.}
\label{Fig:Best_minimun}
\end{figure}

To avoid ending up in local minima for values of $\Pi_0$ differing by $\sim 250$\,s as illustrated for the case in Fig.\,\ref{Fig:grids},  we replaced the actual cost function by its convex approximation and selected the five data points around its minimum. These points were then used to fit a parabola, from which we took the vertex as the final solution for $\Pi_0$. The convex envelope was computed using the \texttt{Qhull} library \citep{convexhull}. The process to obtain the final value of $\Pi_0$ is shown in Fig.\,\ref{Fig:Best_minimun}, where the red and blue vertical lines represent $\Pi_0$ obtained from the discrete dense parameter grid and the parabolic fit, respectively. 
 
\begin{figure}
\centering
\includegraphics[width=\hsize,height=6.3cm,clip]{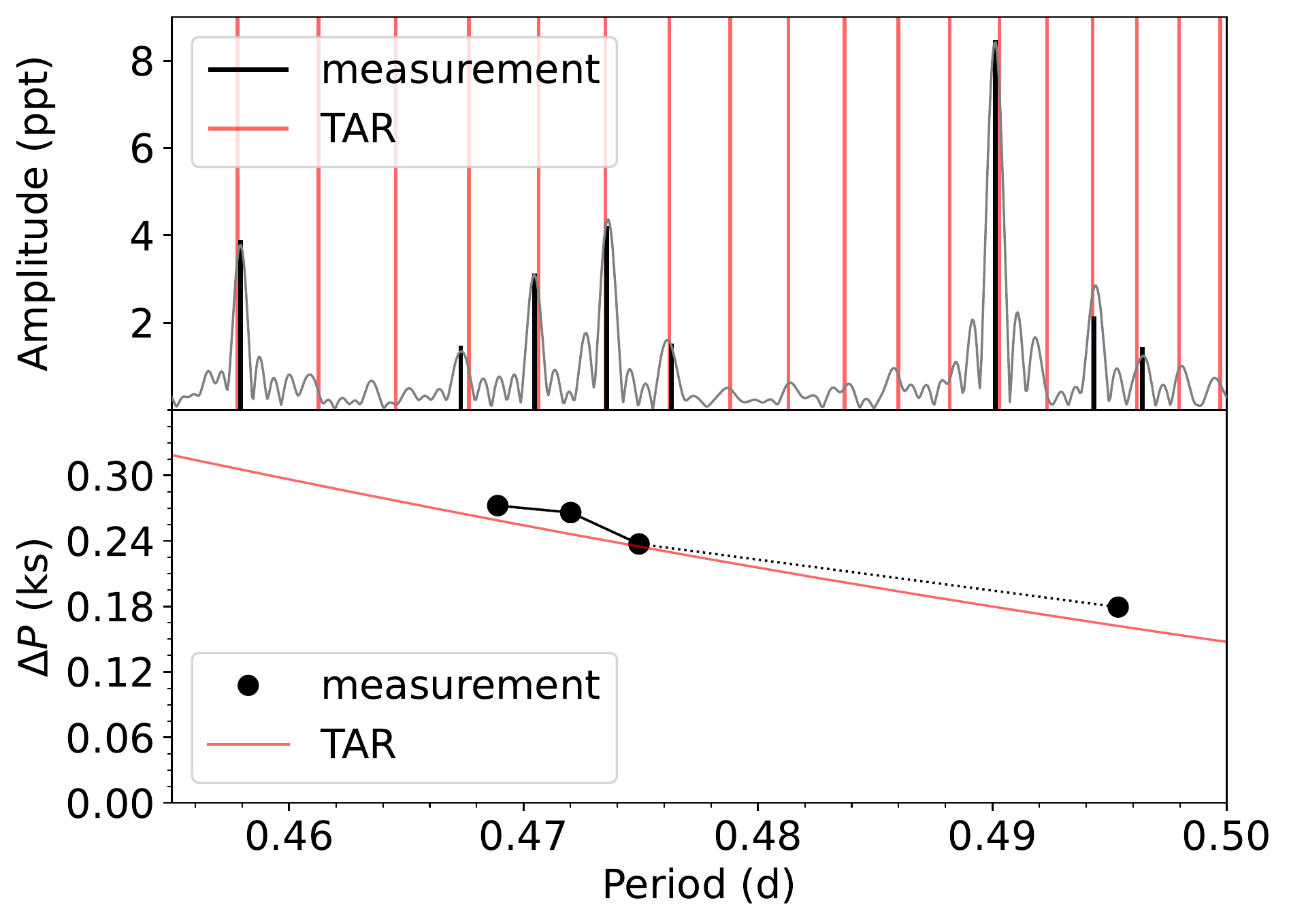}
\caption{Best TAR model for the mode with $k=0$ and $m=1$ in TIC 381950897. Top: Lomb-Scargle periodogram (grey curve), detected periods (black vertical lines), and the TAR fit (red vertical lines). Bottom: Period spacings of both the TAR fit (red curve) and the detected periods (black circles). The latter omits period spacings of missing modes. The dotted line connects periods spacings that stretch across missing modes. The units of the vertical axis $\Delta P$ are  kilosecond (ks) or $10^3$ s.}
\label{Fig:TAR}
\end{figure}

Figure\,\ref{Fig:TAR} shows the final result of the TAR model that best fits the main period-spacing pattern of TIC\,381950897. The lower panel of the figure shows the period spacings in the data where we made the representation such as to exclude the parts with missing modes in the data, although these were identified from the theoretical prediction. Therefore, the lower panel shows a lower or equal number of symbols labelled as measurements when compared to the upper panel, but we stress that the TAR fit was performed using all the detected mode periods indicated in the upper panel. 
 
\subsection{Correlation structure and error estimation}

\begin{figure*}
\resizebox{\hsize}{!}
{\includegraphics[width=\hsize,height=5.5cm,clip]{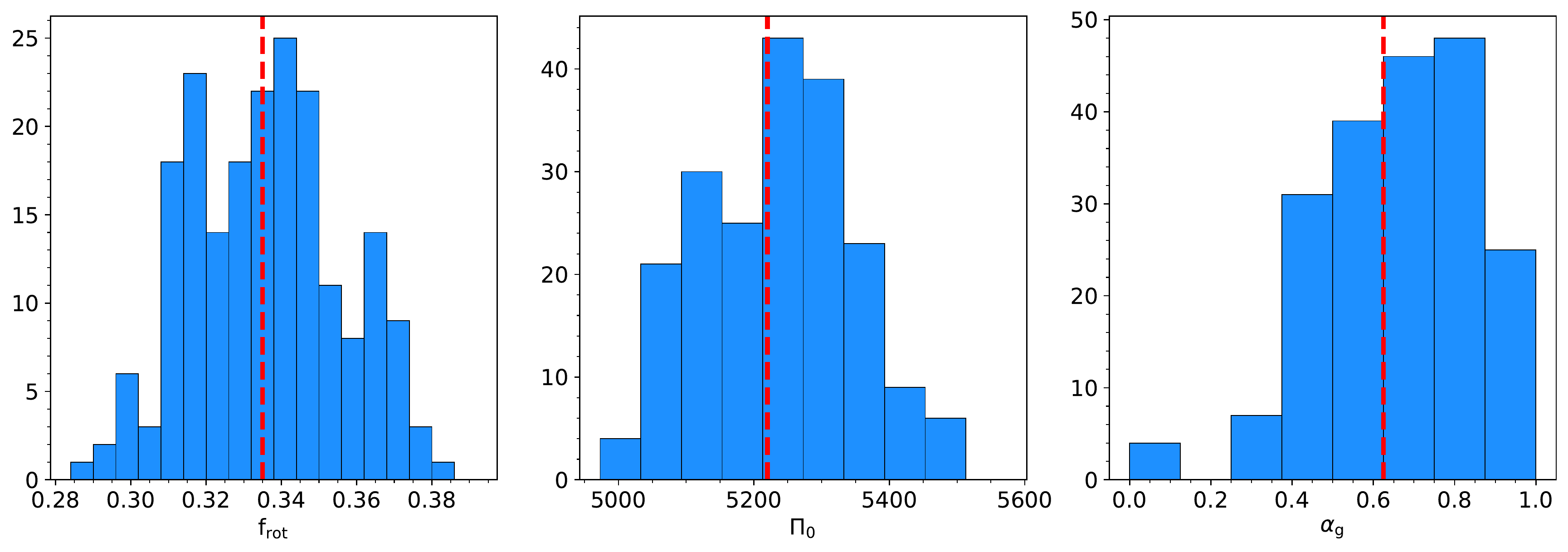}}
\caption{Residual bootstrap distribution for the three fitted parameters of 
TIC\,41483281. The vertical dashed lines indicate the TAR best fit solutions.}
\label{Fig:bootstrap}
\end{figure*}

Correlations between $f_\textrm{rot}$ and $\Pi_0$ found in the parameter space were already presented in Fig.\,\ref{Fig:Pi0_vs_frot}. The shape of the correlation in the left panel is the result of the internal rotation rate affecting the pulsation periods at two separate instances in the methodological chain: 1) from the conversion of the asymptotic spacing for non-rotating values into those for rotating values via the spin parameter; 2) from transformation of the mode periods computed in the co-rotating frame towards the line-of-sight, cf.\,Eq.\,(\ref{Eq:asymptomatic_approx}) and (\ref{Eq:inertial_frame}). When $f_\textrm{rot}$ is such that the two instances affect the pulsation periods similarly, we get a correlation structure as in the right panel of Fig.\,\ref{Fig:Pi0_vs_frot}, as already discussed in the theory paper by \citep{Bouabid2013}.

In order to deal appropriately with the variety of correlation structures, we used a residual bootstrap technique to estimate the uncertainties of $f_\textrm{rot}$ and $\Pi_0$. The errors reported in Table\,\ref{Tab:results} are the $1\sigma$ errors that correspond to the 68\% confidence interval obtained from the bootstrap distributions generated by a non-parametric residual resampling of the TAR fit. The parameter space used during the bootstrap application was centred around the best fit TAR solution and had a full range of 0.12\,d$^{-1}$ for $f_{\rm rot}$ and a range of 1500\,s for $\Pi_0$. These ranges were chosen so as to cover typical uncertainties achieved from 4-yr {\it Kepler\/} light curves based on the median of the $1\sigma$ uncertainties reported in \citet{VanReeth2016}. The sampling for $\alpha_{\rm g}$ for the bootstrap was taken to be the same as the dense parameter space indicated in Table\,\ref{Tab:parameter_space}. 

For each detected pattern, we generated 200 datasets of the same size as the measured one and subsequently minimised Eq.\,(\ref{Eq:cost_function}) for each of them to sample the error distribution and deduce its variance. As an example, Fig.\,\ref{Fig:bootstrap} shows the bootstrap distribution for TIC\,41483281 from which we estimated the $1\sigma$ error for $f_\textrm{rot}$ and $\Pi_0$. We note that the distribution for the parameter $\alpha_g$ is not uniform yet well defined despite the modest resolution we used for this parameter given that it does not offer any astrophysical information but is only a nuisance parameter with the adopted approach (see Table\,\ref{Tab:parameter_space}).

\section{Results for the interior rotation and buoyancy travel time of TESS $\gamma$~Dor stars}\label{Sec:results}
%--------------------------------------

We obtained $f_\textrm{rot}$ and $\Pi_0$ for 60 $\gamma$~Dor stars in the TESS S-CVZ from the catalogue of \citet{Garcia2022} from application of our methodology. For the remaining 46 $\gamma$~Dor stars we did not obtain a converged fit. The patterns were either too short or could not be unambiguously identified because of too many missing modes. Although these 46 stars revealed period-spacing patterns in their TESS 1-yr light curve, they require a longer light curve to achieve mode identification. For the 60 successfully analysed stars, the results are listed in Table\,\ref{Tab:results}, along with information about the mean period, mean period spacing, and the presence of acoustic modes from \citet{Garcia2022}. All the fits to the patterns we obtained are displayed in the supplementary material available online, in the same format as in Fig.\,\ref{Fig:TAR}. A subset of the fits are presented in Sect.\,\ref{Sec:Interesting_stars}. We note that missing modes were common in the catalogue of \citet{Garcia2022} due to the limited duration of the Cycle\,1 S-CVZ TESS data, unlike in the 4-yr long \textit{Kepler\/} light curves.

\begin{figure}
\centering
\includegraphics[width=\hsize,clip]{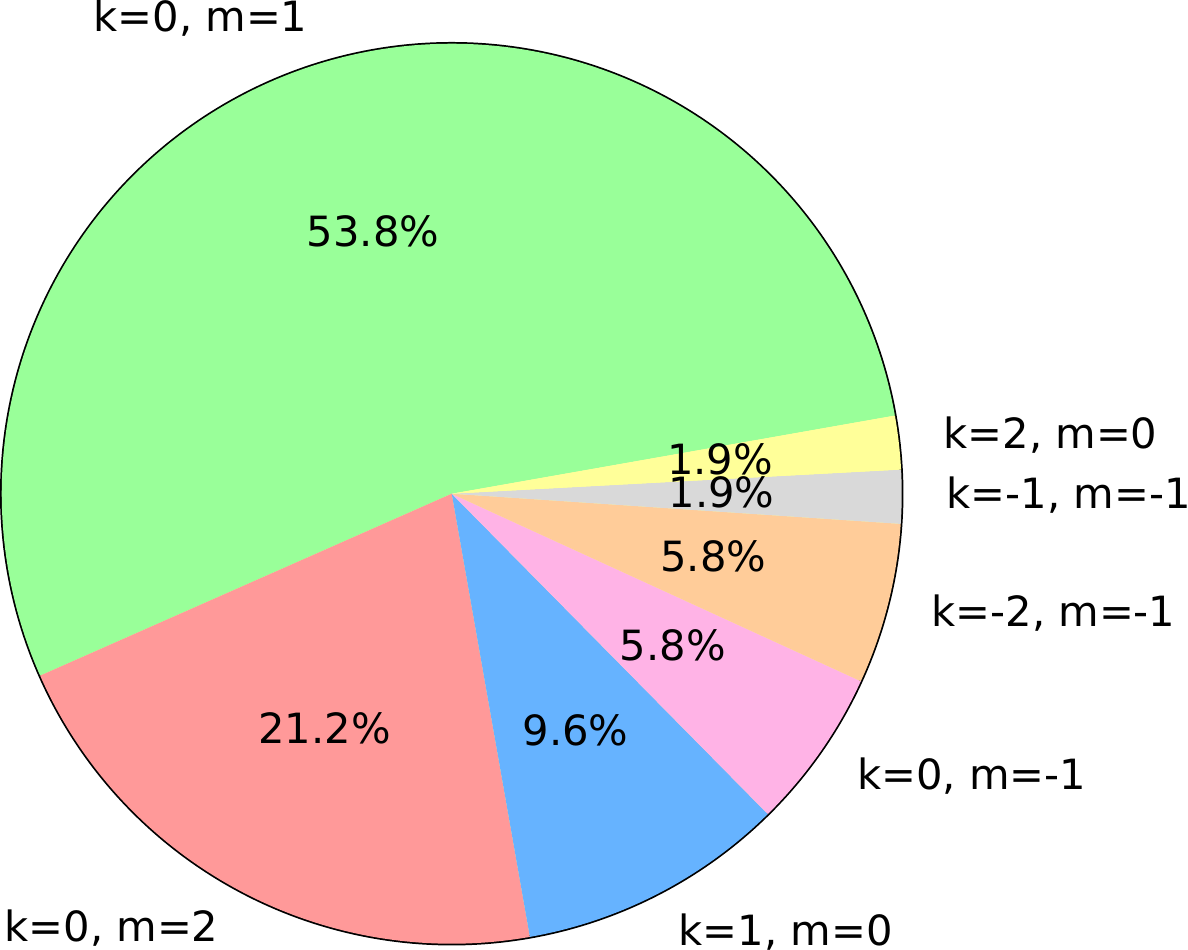}
\caption{Distribution of the identified modes for 60 TESS $\gamma$~Dor pulsators in the catalogue of \citet{Garcia2022}.}
\label{Fig:modeIDs_pie_chart}
\end{figure}

\begin{figure}[h!]
\centering
\includegraphics[width=\hsize,clip,trim={1cm 0 0 0}]{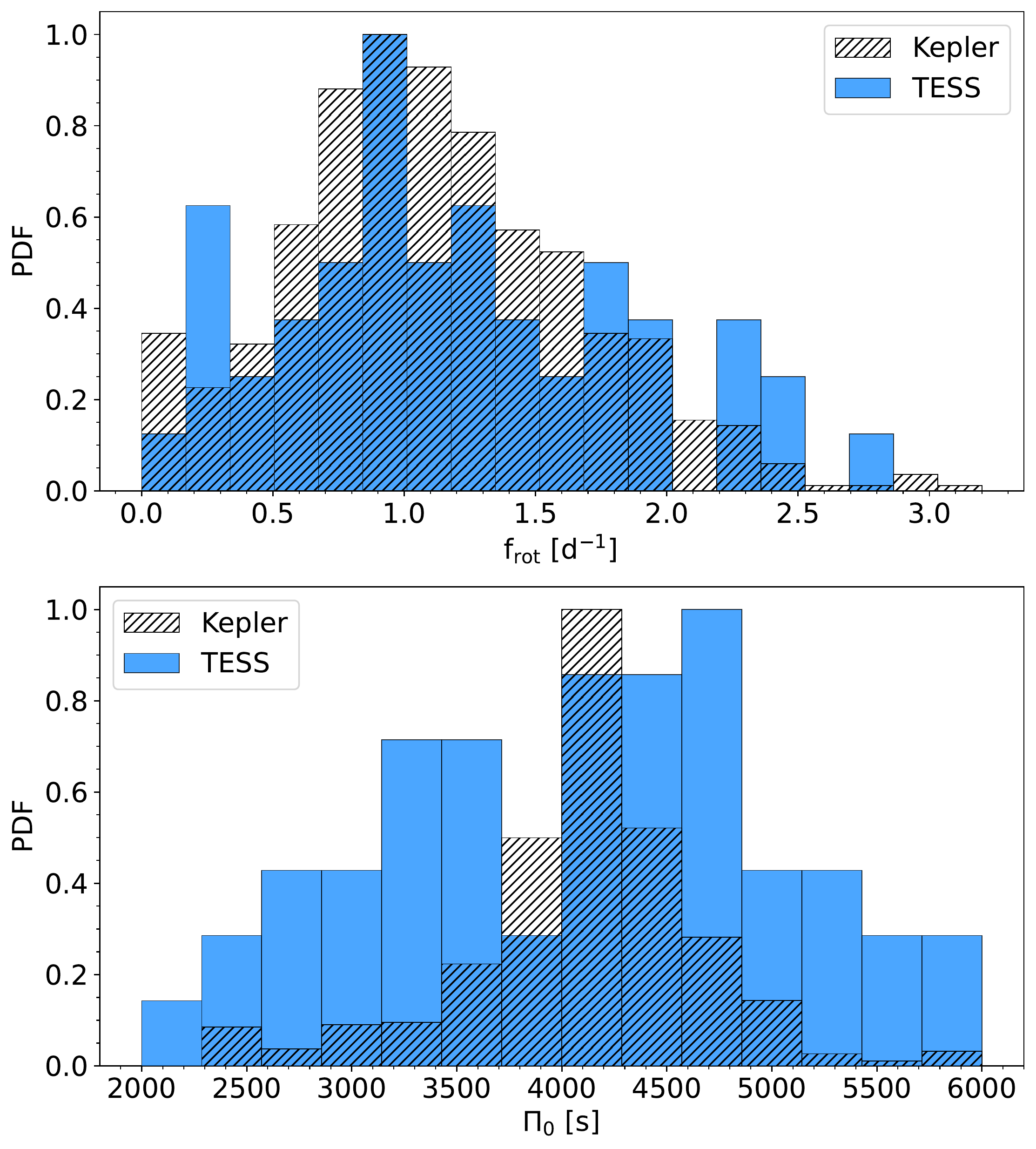}
\caption{Distribution of internal rotation and buoyancy travel time in 611 $\gamma$~Dor stars in the \textit{Kepler\/} sample and the 60 pulsators in our sample from the TESS S-CVZ. The PDF of each histogram has been normalised to unity for comparison.}
\label{Fig:histograms}
\end{figure}

The distribution of identified pulsation modes is presented in Fig.\,\ref{Fig:modeIDs_pie_chart}. It reveals a dominance of dipole $(k=0, m=1)$ and quadruple $(k=0, m=2)$ prograde sectoral modes. This distribution is consistent with the results for $\sim 600$ {\it Kepler\/} $\gamma$~Dor pulsators from \citet{LiGang2020}. The distributions of our values for $f_\textrm{rot}$ and $\Pi_0$ and those for the {\it Kepler\/} sample treated by \citet{LiGang2020} are compared in Fig.\,\ref{Fig:histograms}. Both samples have similar averages despite the factor ten difference in size. 

\begin{figure*}
\resizebox{\hsize}{!}
{\includegraphics[width=\hsize,clip]{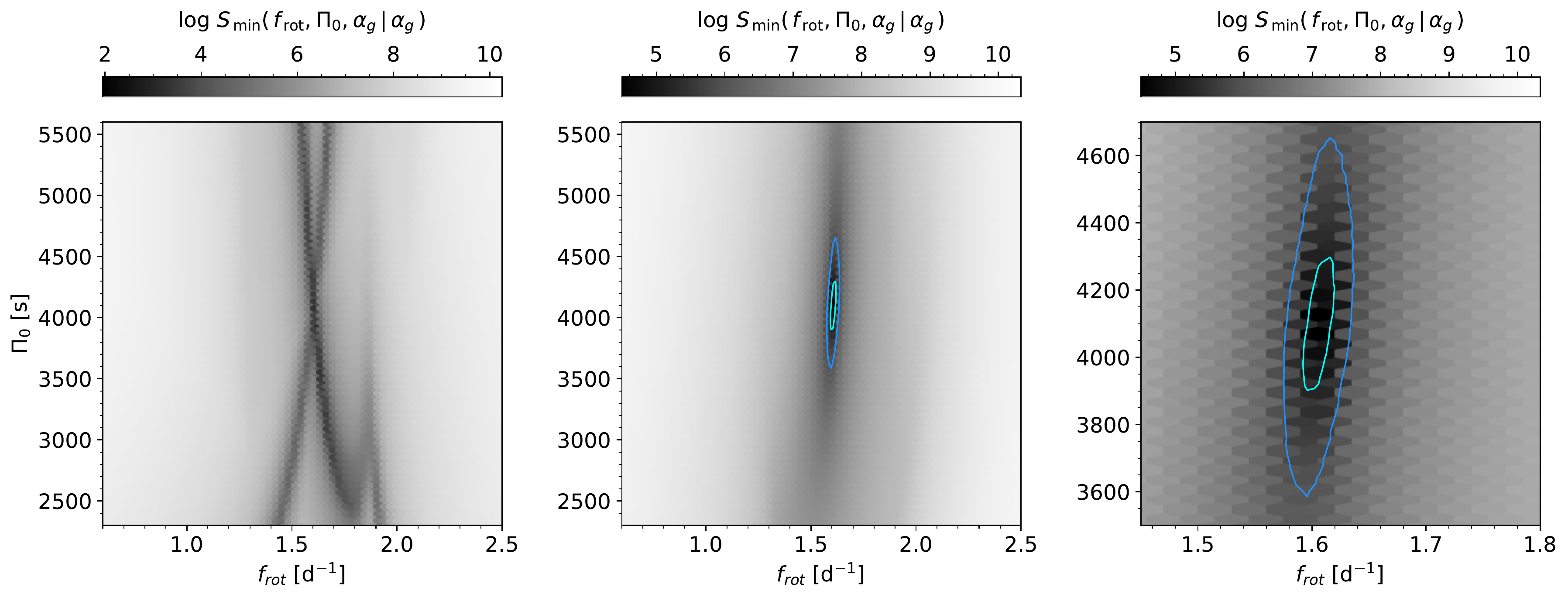}}
\caption{Improvement of the determination of $f_\textrm{rot}$ and $\Pi_0$ for TIC\,381950897 when fitting its two detected period-spacing patterns simultaneously. It concerns a prograde mode (0,1) and a retrograde mode (-2,-1) pattern. The left panel shows an overplot of the two cost functions in which the valley of $S$ with an overall $\mathrm{d}\Pi_0/\mathrm{d}f_\textrm{rot}>0$ corresponds to the prograde mode while the other valley, refers to the retrograde mode. Contour lines are omitted for better visibility but were shown in Fig.\,\ref{Fig:Pi0_vs_frot} where both cost functions are shown separately. The middle panel shows the cost function when both patterns are fitted simultaneously. The solution lies at the intersection of the two valleys from the left panel. The right panel shows a close-up of the middle panel. The blue contour line indicates $S=100\,S_{\rm min}$ while the cyan one indicates $S=10\,S_{\rm min}$.}
\label{Fig:combination}
\end{figure*}

\subsection{Stars with multiple g-mode period-spacing patterns}
%--------------------------------------

Stars with two or more period-spacing patterns provide us with the opportunity to further constrain the stellar interior by fitting the patterns simultaneously \citep[e.g.,][]{VanReeth2016}. \citet{Garcia2022} found 26 of the 106 g-mode pulsators to have  multiple patterns. However, some of the secondary patterns had too few modes to obtain a converged fit. We managed to get converged simultaneous fits for 5 of the 26 stars, as listed in Table\,\ref{Tab:results2}. We relied on the pattern with the clearest mode identification to resolve any uncertainty in the identification of the secondary pattern, following some criteria: (1) modes with  $S_{\rm min}^{\rm global}/S_{\rm min}^{\rm km}<0.25$ were rejected, see Fig. \ref{Fig:mode_id} for an example, (2) the two independent solutions for $\Pi_0$ must be consistent within $2\sigma$, (3) solutions based on equality of the two independent $f_{\rm rot}$ are preferred, given that almost all {\it Kepler\/} $\gamma$~Dor stars reveal near-rigid rotation \citep{LiGang2020}, (4) prograde sectoral modes are preferred following the mode identification distribution obtained by \citet{LiGang2020}.

An example of a simultaneous two-pattern fit is shown in Fig.\,\ref{Fig:combination} for TIC\,381950897. The left panel shows the cost function for $f_\textrm{rot}$ and $\Pi_0$ when the two patterns are fitted independently. The middle and right panels show the better constrained solution when the two patterns are fitted simultaneously. The minimum of the cost function is indeed more localised, as shown by the comparison between the contours in Figs\,\ref{Fig:Pi0_vs_frot} and \ref{Fig:combination}. This translates into more precise values for $f_\textrm{rot}$ and $\Pi_0$, as indicated in 
Table\,\ref{Tab:results2}.

\subsection{Most promising TESS $\gamma$~Dor stars for future asteroseismic modelling}\label{Sec:Interesting_stars}

\begin{figure*}
\resizebox{\hsize}{!}
{\includegraphics[width=\hsize,clip]{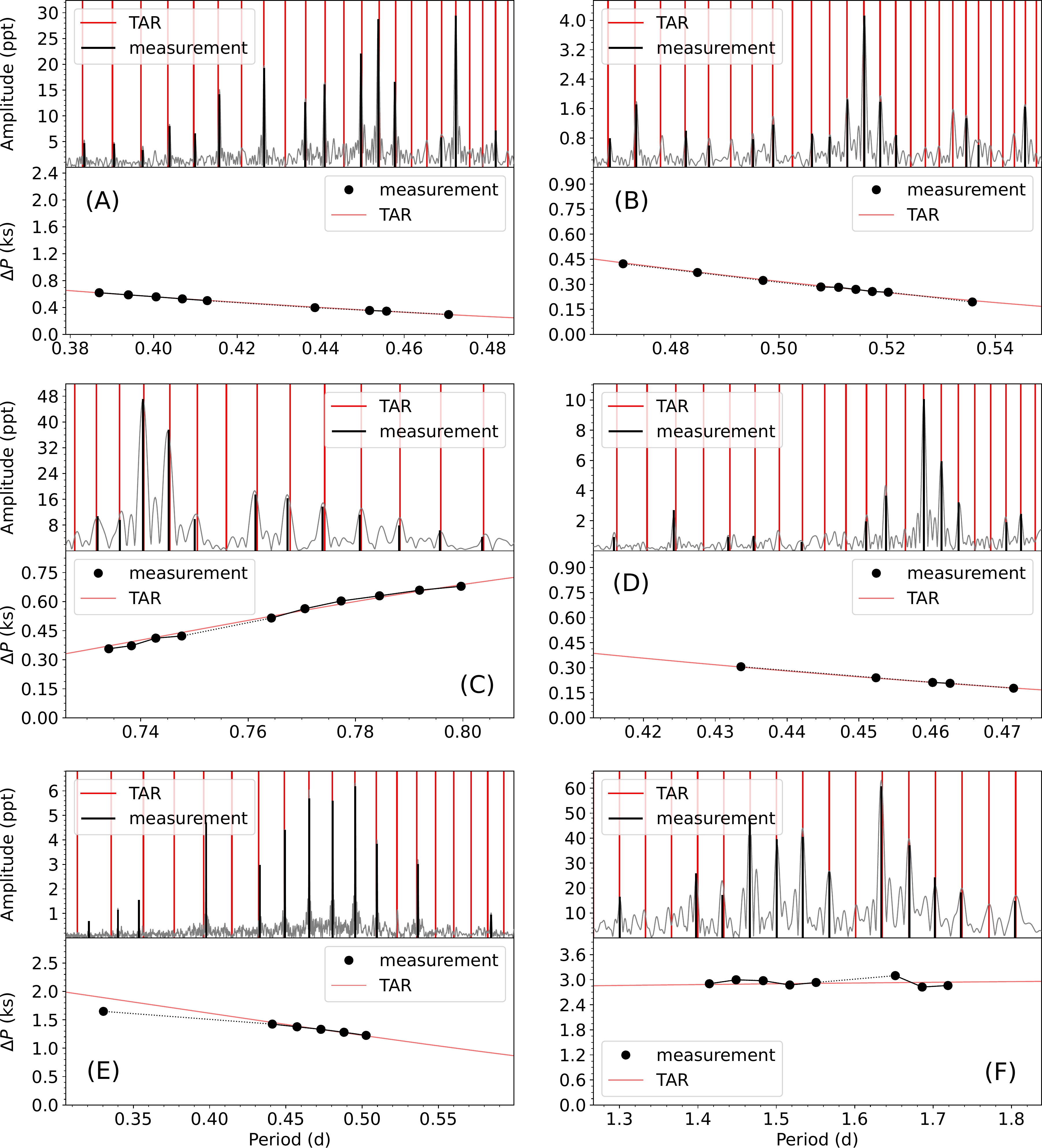}}
\caption{Most promising TESS $\gamma$~Dor stars for asteroseismic modelling, revealing the best TAR fits. (A): TIC\,374944608. (B): TIC\,350144657. (C): TIC\,381950897. (D): TIC\,349832567. (E): TIC\,141479660. (F): TIC\,38515566. Symbols are the same as in Fig.\,\ref{Fig:TAR}.}
\label{Fig:fits}
\end{figure*}

Six stars in our sample offer particularly long patterns with few missing modes and hence have good potential for future asteroseismic modelling. These are TIC\,374944608, TIC\,350144657, TIC\,381950897, TIC\,349832567, TIC\,141479660, and TIC\,38515566. These six stars show period-spacing patterns with 12 to 15 frequencies with high amplitudes and signal-noise ratio in the Lomb-Scargle periodograms, as presented in Fig.\,\ref{Fig:fits}. Most of these stars are also hybrid pulsators, that is they reveal both g and p (acoustic) modes. They also have available spectroscopic data as discussed in the next section (and listed in Table\,\ref{Tab:results}). This makes these stars prime candidates for further asteroseismic modelling as in \citet{Mombarg2020,Mombarg2021}.

Adding the TESS Cycle 3 and 5 data to the light curves of these stars (as all others in our sample) will translate into effectively three and 5-yr long light curves, respectively. Although those light curves will have one/two 1-yr gaps, they will undoubtedly decrease the noise levels in the periodograms, while still delivering manageable aliasing. This will allow us to detect some of the currently missing modes and prolong the period-spacing patterns found here, in addition to increasing frequency resolution appreciably. 

\section{Astrophysical properties of the TESS $\gamma$~Dor stars}\label{Sec:spec_characterization}

\begin{figure*}
\resizebox{\hsize}{!}
{\includegraphics[width=\hsize,clip]{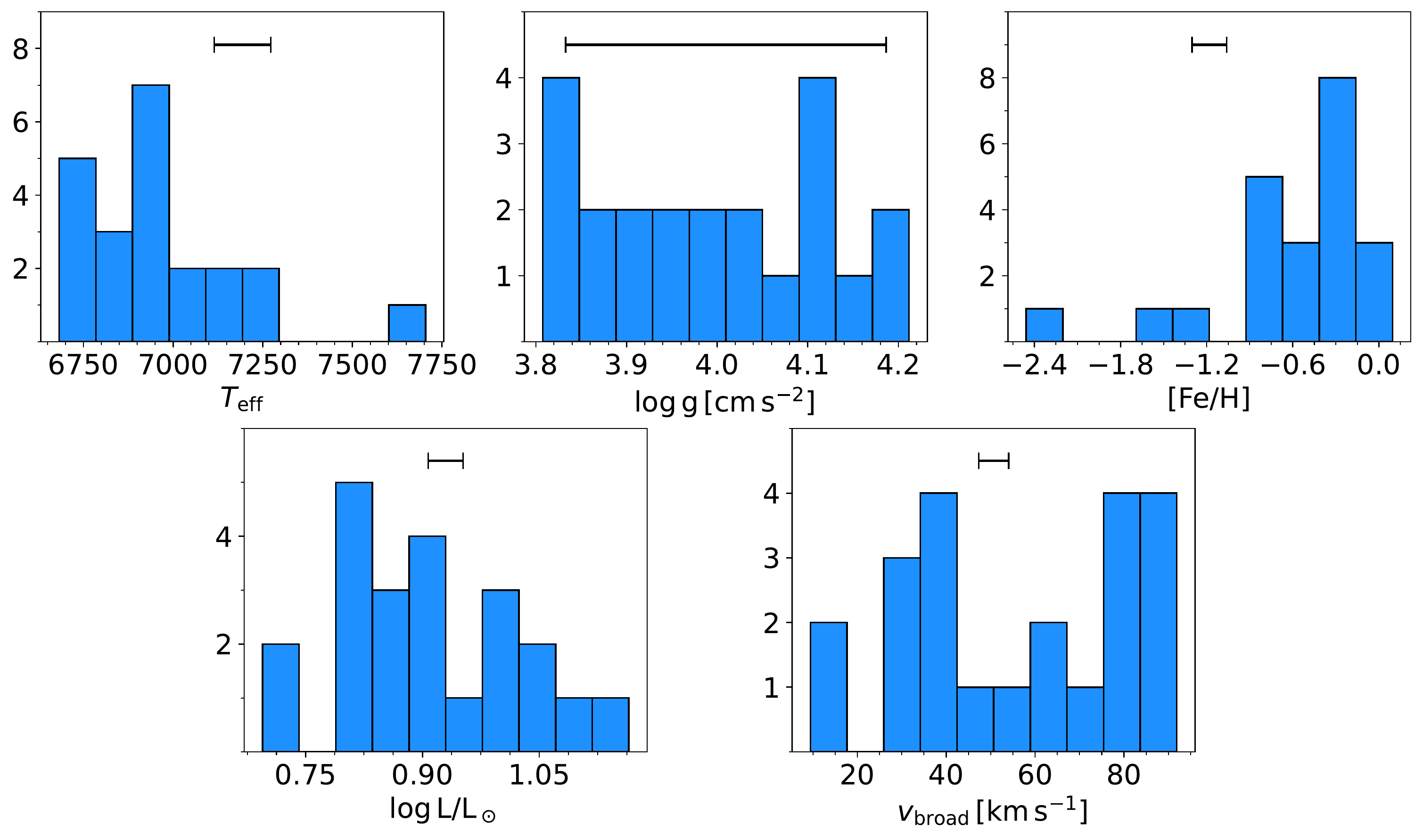}}
\caption{Distribution of spectroscopic parameters and luminosity for 22 stars in our catalogue. Typical uncertainties are shown at the top of each panel. See Section \ref{Sec:spec_characterization} for a further description.}
\label{Fig:spec_characterization}
\end{figure*}

We searched the literature for homogeneous spectroscopic parameters of our sample stars and found 34 of them to have spectra available in the Third Data Release (DR3) of the GALAH Survey \citep{Buder2021}. For 22 of them, reliable stellar parameters were determined\footnote{Flagged with \texttt{flag\_sp==0} in the GALAH database.}.

We took the spectroscopic parameters effective temperature, gravity, metallicity, and the velocity connected with the spectral line broadening, $v_{\rm broad}$, which is an upper limit of the projected rotation velocity $v\sin i$ in the analysis method adopted by \citet{Buder2021}. Indeed, the authors fitted the broadened metal lines in the medium-resolution GALAH spectra assuming the absence of any macroturbulence, often connected with the occurrence of pulsations, notably g modes \citep{Aerts2009}. We took the luminosity for these 22 $\gamma$~Dor stars from the TESS Input Catalog, Version 8 \citep[based on Gaia DR2 parallaxes]{Stassun2019}. The distributions of the stellar parameters for these 22 $\gamma$~Dor stars are shown in Fig.\,\ref{Fig:spec_characterization}. These distributions for the effective temperature and surface gravity are consistent with those of {\it Kepler\/} $\gamma$~Dor stars from high-resolution spectroscopy \citep{Niemczura2015,VanReeth2015b,Gebruers2021}. However, the metallicities reported in the GALAH study by \citet{Buder2021} are lower than those found by \citet{Gebruers2021} based on much higher quality spectroscopic data, which may point to major  systematic uncertainties in this quantity deduced for F-type dwarfs from medium-resolution survey spectroscopy. The distribution for $v_{\rm broad}$, when compared with the rotational broadening for {\it Kepler\/} $\gamma$~Dor stars derived by \citet[][their Fig.\,5]{Gebruers2021}, shows that these 22 TESS $\gamma$~Dor stars with identified patterns are mainly slow rotators. 

\begin{figure}[h!]
\centering
\includegraphics[width=\hsize,clip]{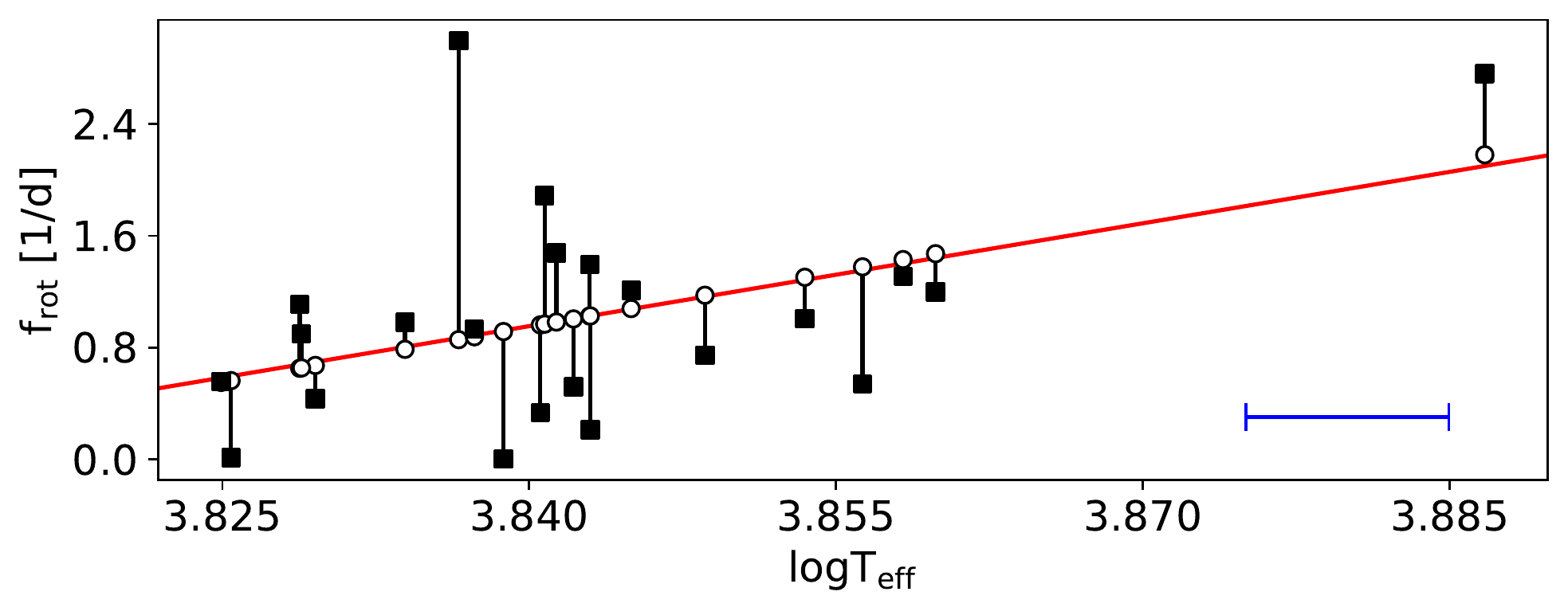}
\caption{Multivariate linear regression for the internal rotation. The squares are the observations and the circles are the modelled values. The corresponding values are connected with a black line. The red line is the model listed in Table \ref{Tab:MLR_frot}. The typical uncertainty of ${\rm log}\, T_{\rm eff}$ is shown in blue at the bottom-right corner. Uncertainties of $f_{\rm rot}$ are smaller than the symbol sizes.}
\label{Fig:MLR_frot}
\end{figure}

\begin{figure}[h!]
\centering
\includegraphics[width=\hsize,clip]{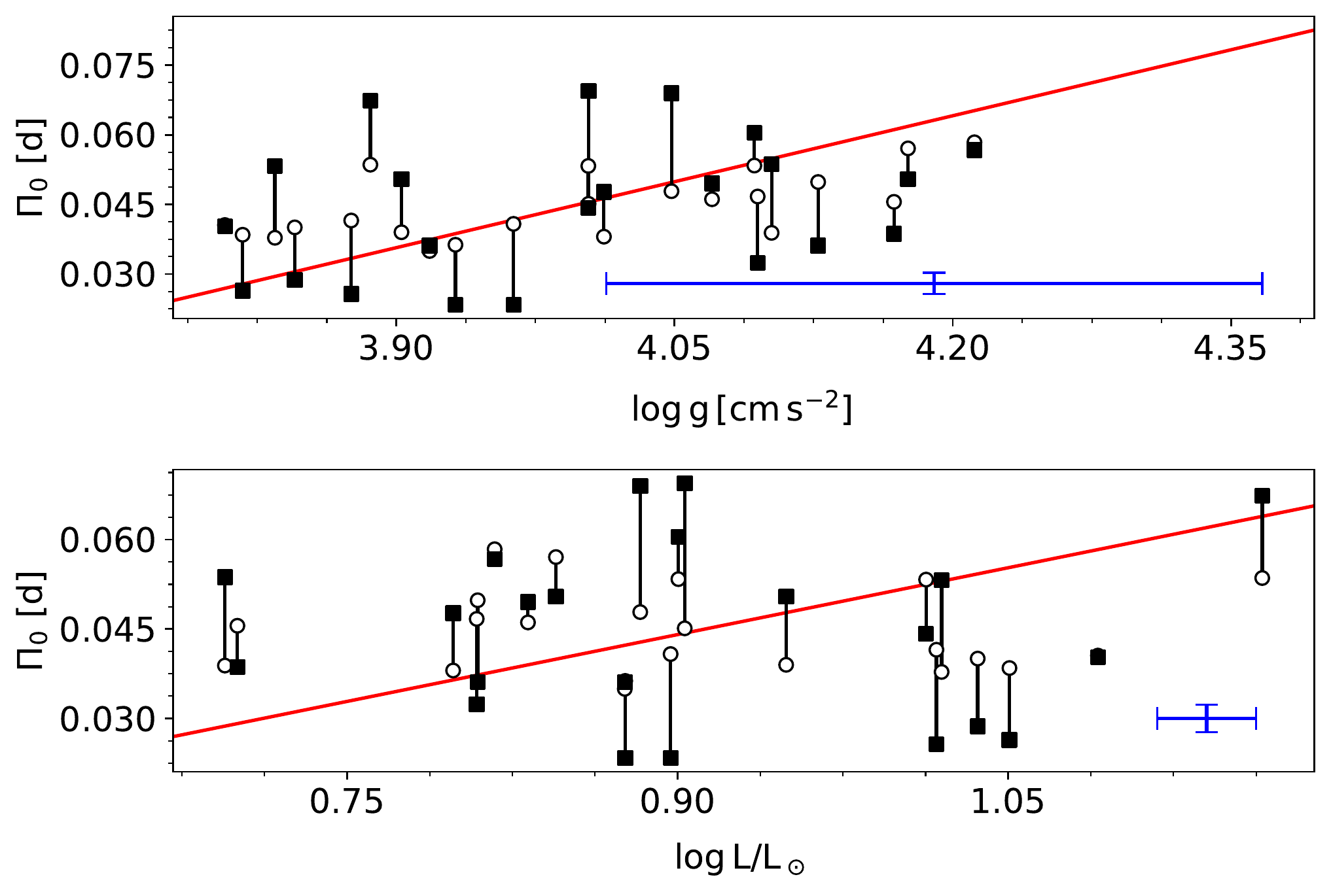}
\caption{Multivariate linear regression for the buoyancy travel time. Each panel shows The bivariate model fits listed in Table \ref{Tab:MLR_Pi0}. The symbols are the same as in Fig. \ref{Fig:MLR_frot}. Typical uncertainties are shown in blue at the bottom-right corner.}
\label{Fig:MLR_Pi0}
\end{figure}

Despite the small sample of only 22 $\gamma$~Dor stars, we searched for correlations among $f_{\rm rot}$, $\Pi_0$ and the stellar parameters shown in Fig.\,\ref{Fig:spec_characterization}. Following \citet{VanReeth2015b}, we used multivariate linear regression with backwards elimination, that is each of the 
rotation frequencies and buoyancy travel times were modelled as a linear combination of the five stellar parameters shown in Fig.\,\ref{Fig:spec_characterization}. The coefficient with the largest p-value above 0.05, obtained from a $t$-test, was then removed from the regression model before repeating it, until only explanatory variables with significant p-values remained among the stellar parameters. We found a weak correlation ($R^2=0.20$) between $f_{\rm rot}$ and $T_{\rm eff}$ as shown in Fig.\,\ref{Fig:MLR_frot} and whose free parameters are listed in Table\,\ref{Tab:MLR_frot}. Despite the small $R^2$ value, this correlation may indicate that stars on the main sequence slow down as they age, as expected from angular momentum transport mechanisms active during stellar evolution \citep{Aerts2019}. In contrast, no significant correlation was found between $f_{\rm rot}$ and $v_{\rm broad}$, adding evidence for the presence of macroturbulence affecting the line broadening. We also found a weak bivariate relationship ($R^2=0.23$) between $\Pi_0$ on the one hand, and ${\rm log}\,g$ and ${\rm log}\,L$ on the other hand. The positive parameters of the bivariate fit listed in Table\,\ref{Tab:MLR_Pi0} and shown in Fig.\,\ref{Fig:MLR_Pi0} indicate that the buoyancy travel time increases as the luminosity or surface gravity increase. \citet{VanReeth2016} found similar weak correlations expected from stellar evolution theory for the rotation, yet none for the buoyancy travel time. 

\begin{figure}[h!]
\centering
\includegraphics[width=\hsize,clip]{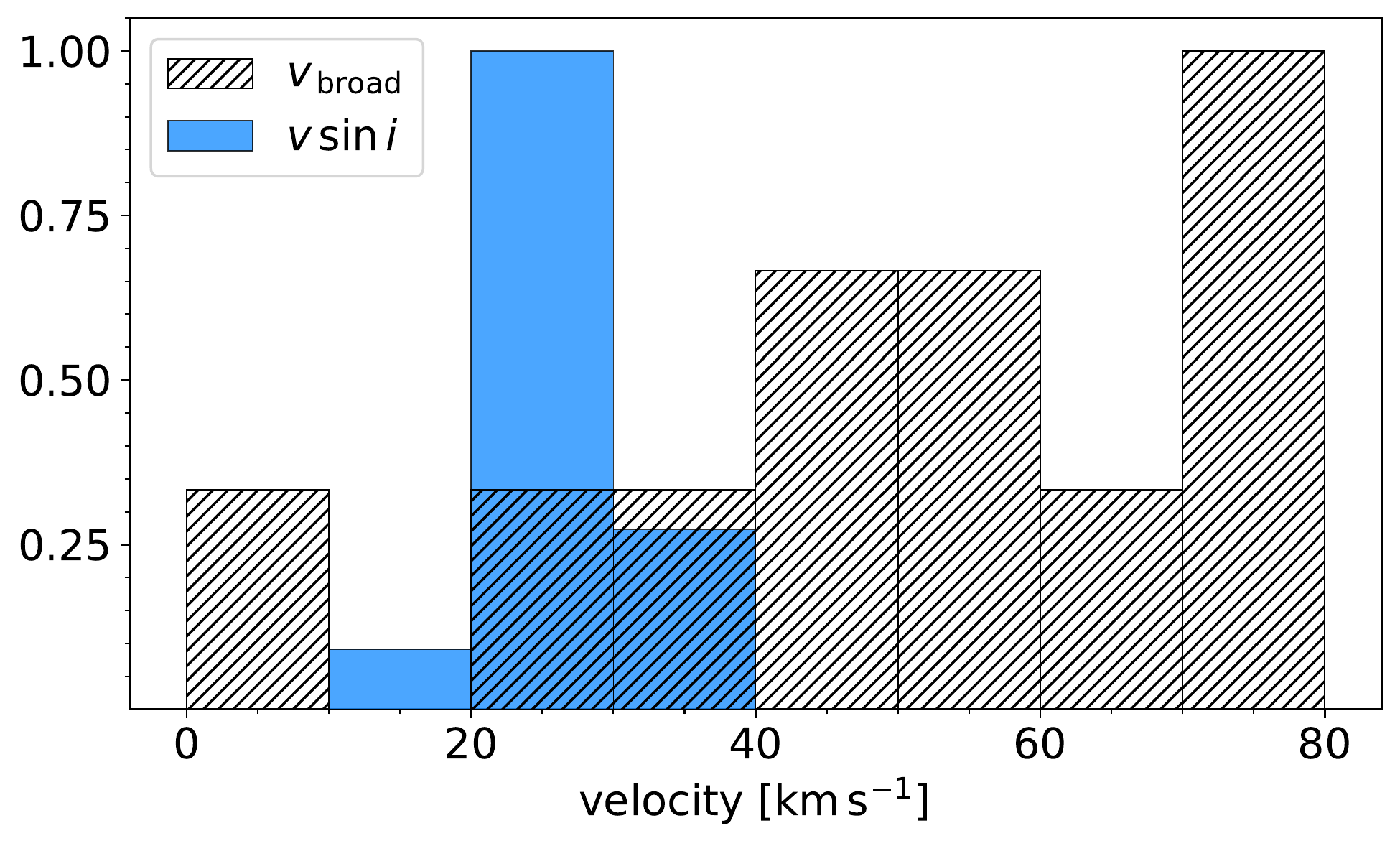}
\caption{Broadening velocities and projected rotational velocities of our 15 stars in common with GALAH DR3 with bona fide spectra and with reported values from \cite{Sharma2018}. Histograms have been normalised to unity for a comparison.}
\label{Fig:Vbroadening_and_vsini}
\end{figure}

\begin{figure*}
\resizebox{\hsize}{!}
{\includegraphics[width=\hsize,clip]{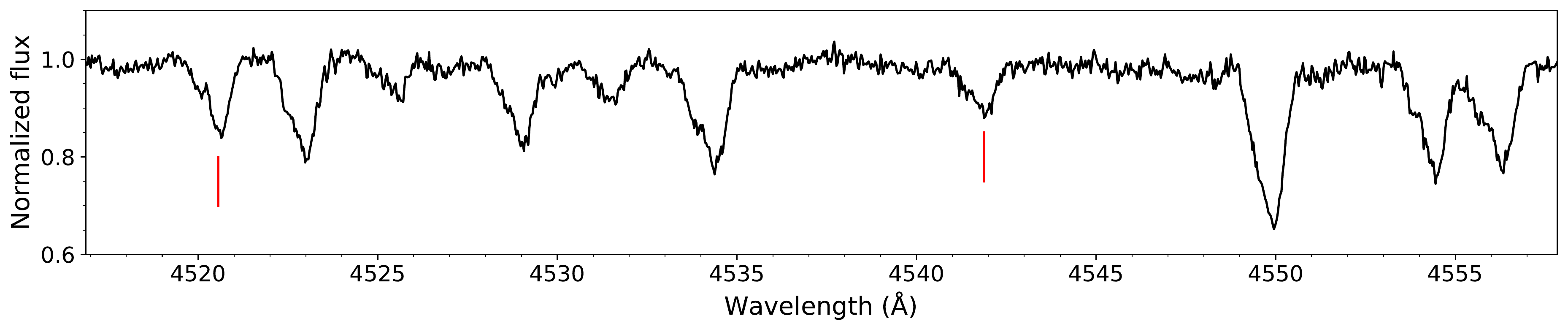}}
\caption{Normalised spectrum of TIC 350343297 showing asymmetric absorption lines. The lines FeII$\lambda$4520.2 and FeII$\lambda$4541.5 are indicated with red vertical lines as a reference. Spectrum obtained by the FEROS spectrograph \citep{Kaufer1997,Kaufer1999} at the MPG/ESO 2.2-metre telescope located at the La Silla Observatory in Chile, on date 2021-03-24 with a spectral resolution of 48000 and a signal-to-noise ratio of 104. The spectrum is not corrected by redshift.}
\label{Fig:asymmetry}
\end{figure*}

\begin{figure*}
\resizebox{\hsize}{!}
{\includegraphics[width=\hsize,clip]{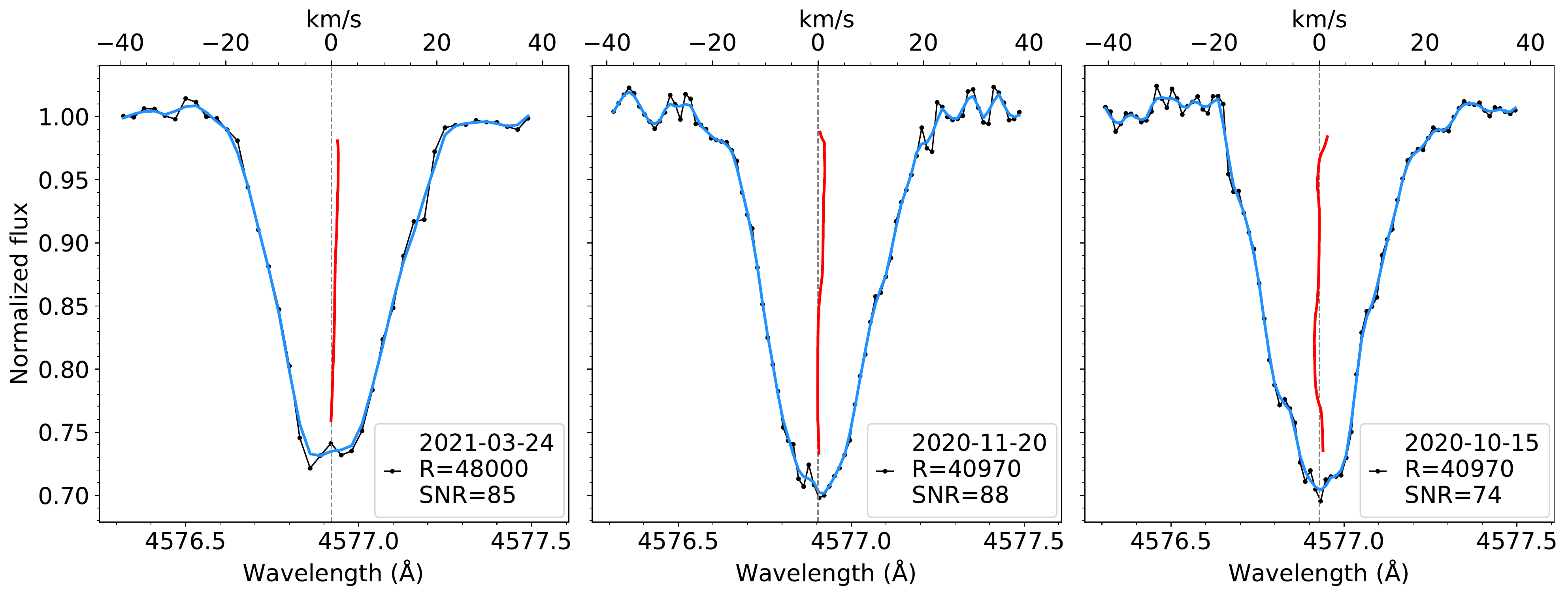}}
\caption{FeII$\lambda$4576.3 absorption line of TIC 167124706 at three different epochs. The bisector of the line is shown in red and is calculated with respect to the smoothed spectrum shown in blue. The smoothed spectrum is generated by a moving average using window length of three data points. The top axes shows the Doppler velocity with respect to the central wavelength of the absorption line indicated with the vertical dashed line. The spectrum is not corrected by redshift.}
\label{Fig:asymmetry_epochs}
\end{figure*}

Values for the projected rotational velocity, $v\sin i$, were available in SIMBAD for 15 of the 22 $\gamma$~Dor stars in the GALAH survey. These values come from \cite{Sharma2018}. We noticed that the distribution of $v\sin i$ from SIMBAD differs from the one of the GALAH $v_{\rm broad}$ values, as shown in Fig.\,\ref{Fig:Vbroadening_and_vsini}. This indicates that time-dependent pulsational line broadening is active in these stars as expected for g modes. Considerable line-profile broadening in addition to  time-independent rotational broadening is indeed a well-known phenomenon detected and analysed for bright $\gamma$~Dor pulsators studied with high-resolution spectroscopy \citep{DeCat2006}. Time-dependent pulsational line broadening is often approximated by time-independent macroturbulence and its occurrence may prevent proper derivation of $v\sin i$ \citep{Aerts2014}, which may explain the difference in distributions in Fig.\,\ref{Fig:Vbroadening_and_vsini}. In order to investigate this further, we searched for evidence of line asymmetries in narrow metal lines of the $\gamma$~Dor stars in our sample. We looked for optimal isolated iron spectral lines following the analyses by \citet{Bruntt2008} and computed their bisector as a good diagnostic for line asymmetries. Only 8 of the 15 stars have spectra of high-resolution, namely $R \gtrsim 4800$, with a signal-noise ratio ${\rm SNR}\gtrsim 100$. We found indications of asymmetric lines in the following eight stars: TIC 38515566, TIC 350343297, TIC 349092320, TIC 55849446, TIC 382519218, TIC 149540525, TIC 167124706 and, TIC 150392753. Some of those stars, like TIC 350343297 in Fig.\,\ref{Fig:asymmetry}, show clear asymmetric lines by visual inspection, while others are less pronounced but detected in their bisectors and the change over different epochs, such as  TIC 167124706 in Fig.\,\ref{Fig:asymmetry_epochs}.

\renewcommand{\arraystretch}{1.2}

\begin{table}[h]
\centering
\caption{Multivariate linear regression result for the rotation frequency.}
\begin{tabular}{cccc}
 \hline \hline
coefficient        & value & uncertainty & p-value \\ \hline \hline
const              & -93.3 & 42.1        & 0.039   \\ \hline
${\rm log}\,T_{\rm eff}$ & 24.5  & 10.9        & 0.037  \\ \hline
\end{tabular}
\tablefoot{$R^2=0.20$. Plot shown in Fig.\,\ref{Fig:MLR_frot}.}
\label{Tab:MLR_frot}
\end{table}

\begin{table}[h]
\centering
\caption{Multivariate linear regression for the buoyancy travel time.}
\begin{tabular}{cccc}
 \hline \hline
coefficient        & value & uncertainty & p-value \\ \hline \hline
const              & -0.4 & 0.1        & 0.050   \\ \hline
${\rm log}\,g$ & 0.094  & 0.040        & 0.029  \\ \hline
${\rm log}\,L$ &  0.074  &  0.041        & 0.084  \\ \hline
\end{tabular}
\tablefoot{$R^2=0.23$. Plot shown in Fig.\,\ref{Fig:MLR_Pi0}.}
\label{Tab:MLR_Pi0}
\end{table}

\section{Quality comparison by revisiting {\it Kepler\/} $\gamma$~Dor stars
\label{Sec:Kepler_stars}}

To validate our methodology described in Section\,\ref{Sec:method}, we applied it to {\it Kepler\/} 4-yr light curves of $\gamma$~Dor stars whose $f_{\rm rot}$ and $\Pi_0$ have been derived by \citet{VanReeth2016}. Their sample includes 37 stars that have been monitored with high-resolution spectroscopy by \citet{Tkachenko2013} and were confirmed to be single $\gamma$~Dor stars. The mass, age, and level of near-core mixing of these stars has been deduced from forward asteroseismic modelling by \citet{Mombarg2019,Mombarg2021} so this sample constitutes the best characterised $\gamma$~Dor stars to date. We revisited their {\it Kepler\/} light curve and deduced the g-mode frequencies following the same analysis as in \citet{Garcia2022}.

\begin{figure*}
\resizebox{\hsize}{!}
{\includegraphics[width=\hsize,clip]{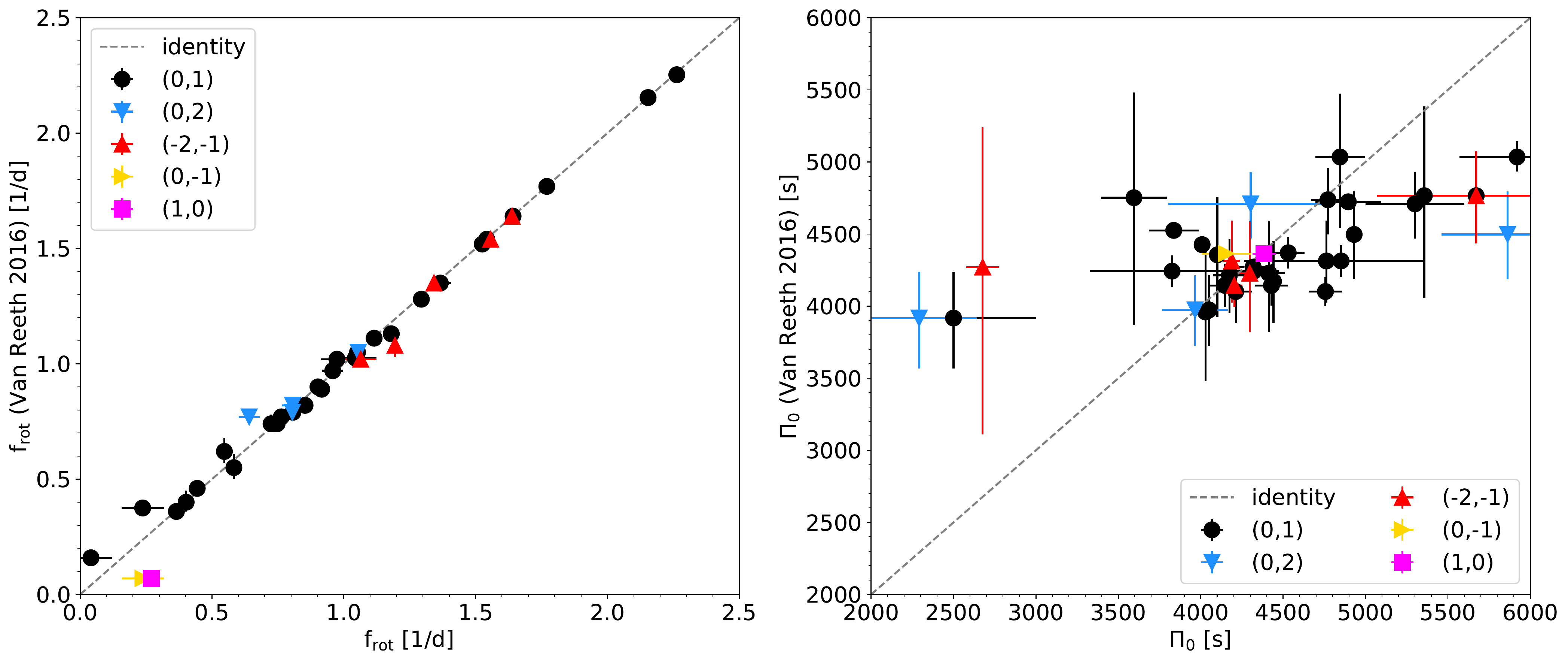}}
\caption{Comparison of internal rotation and buoyancy travel time obtained by \citet{VanReeth2016} and this work. Different colours/symbols are different pulsation modes $(k,m)$ specified in the legend. The symbols labelled with KIC numbers in the left panel are the same patterns labelled in the right panel.}
\label{Fig:comparison_with_VanReeth}
\end{figure*}

\begin{figure*}
\resizebox{\hsize}{!}
{\includegraphics[width=\hsize,clip]{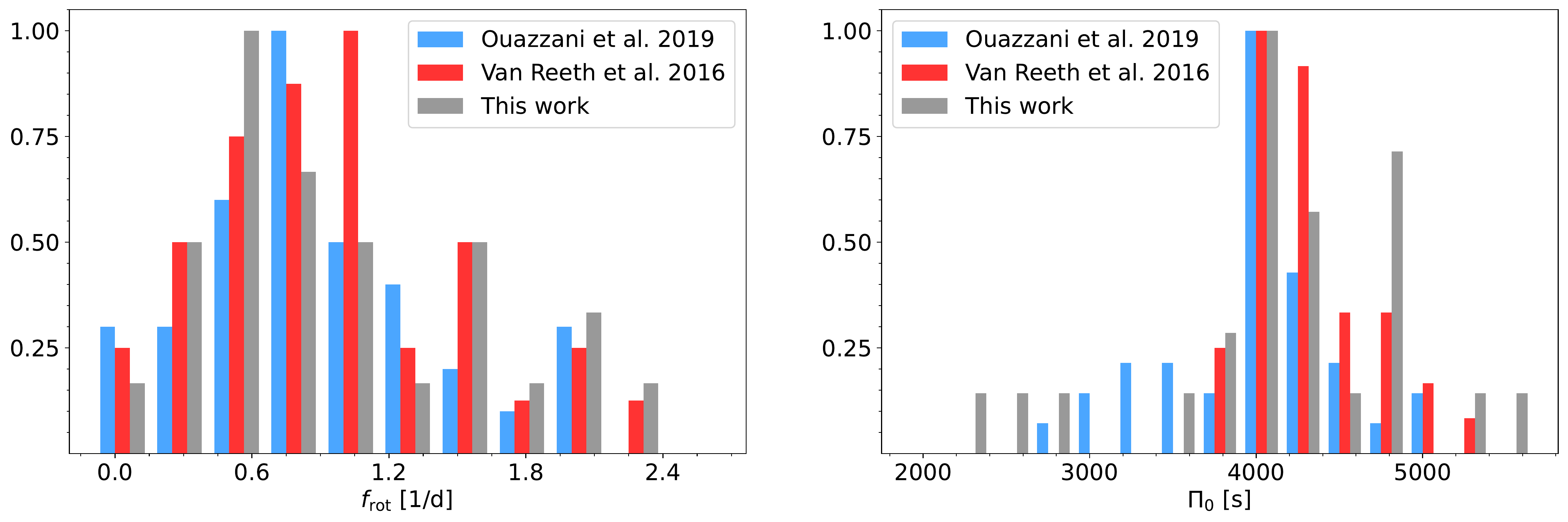}}
\caption{Comparison of internal rotation and buoyancy travel time obtained by \citet{VanReeth2016}, \citet{Ouazzani2019} and this work. Histograms have been normalised to unity for comparison. The columns of different studies are shifted for visibility instead of stacked.}
\label{Fig:comparison_with_VanReeth_and_Ouazzani}
\end{figure*}

\begin{figure}
\resizebox{\hsize}{!}
{\includegraphics[width=\hsize,clip]{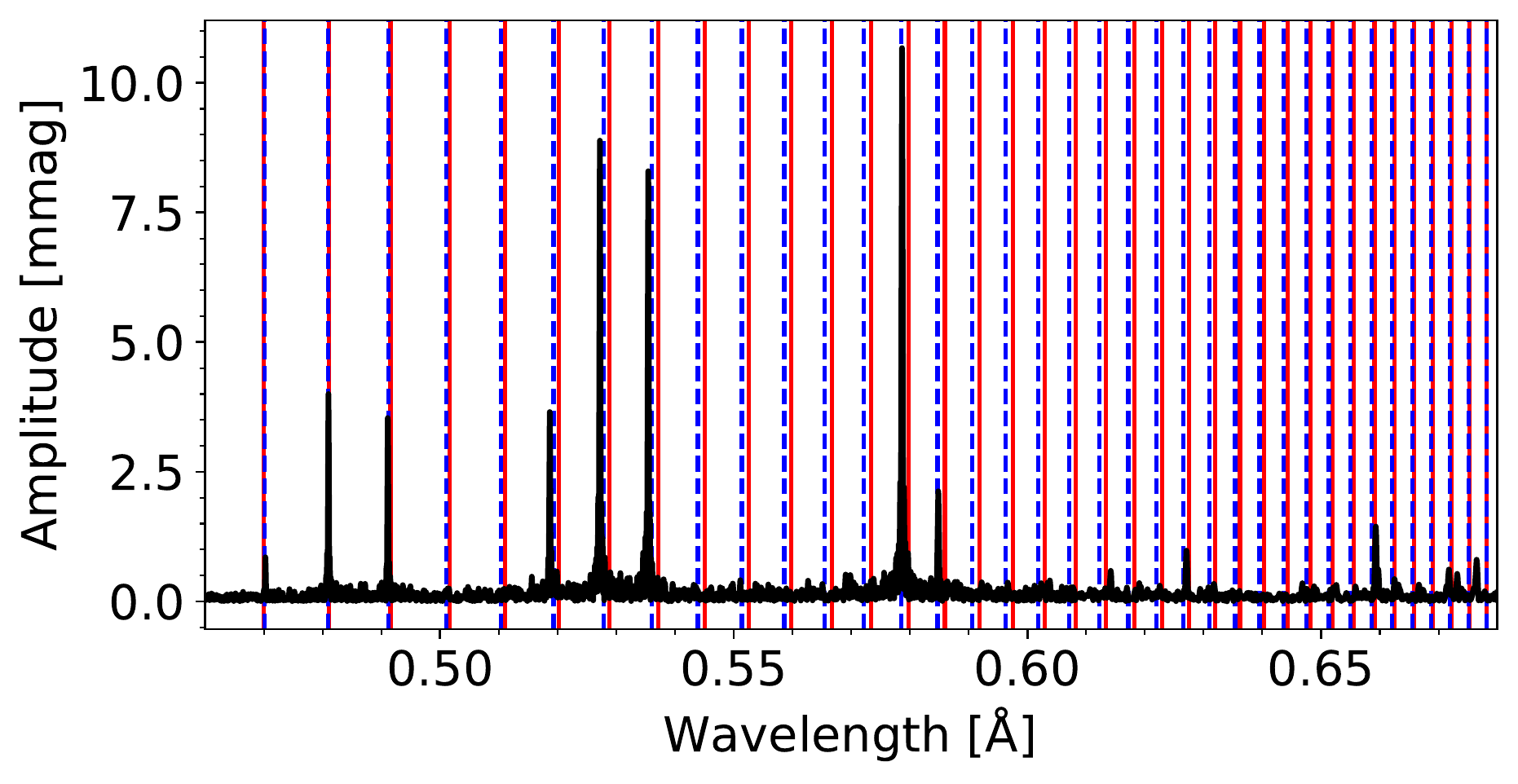}}
\caption{Comparison of the best TAR model obtained by \citet{VanReeth2016} (dashed blue lines) and this work (solid red lines) for the period-spacing pattern in KIC 7939065 reported by the aforementioned study.}
\label{Fig:comparison_with_VanReeth_TAR}
\end{figure}

\citet{VanReeth2016} identified multiple period-spacing patterns in the 37 {\it Kepler\/} stars, but only the patterns for consecutive radial-order modes with $(k,m)=(0,1)$ were used to constrain  $f_{\rm rot}$ and $\Pi_0$. We recovered and identified all patterns as $(k=0,m=1)$. Figure\,\ref{Fig:comparison_with_VanReeth} shows the comparison between our results and those by \citet{VanReeth2016}. The left panel shows excellent agreement for $f_{\rm rot}$. Previously, \citet[][their Fig.\,5]{Ouazzani2019} had already made a comparison of their 37 $f_{\rm rot}$ values deduced by \citet{Christophe2018} independently of \citet{VanReeth2016}. They found the same $f_{\rm rot}$ values than \citet{VanReeth2016} for 36 of the 37 stars. The right panel of Fig.\ref{Fig:comparison_with_VanReeth} compares our $\Pi_0$ values with those deduced by \citet{VanReeth2016}. Overall, this figure  also shows a good agreement for $\Pi_0$ because the larger deviations from the bisector are accompanied by larger uncertainties than for $f_{\rm rot}$. Also this quantity for these 37 stars was deduced independently by \citet{Christophe2018}. Figure\,\ref{Fig:comparison_with_VanReeth_and_Ouazzani} compares the values from the three observational studies starting from the same light curves of the 37 pulsators. While there is overall excellent agreement, particularly for $f_{\rm rot}$, \citet{Ouazzani2019} and this work both find a larger range of values for $\Pi_0$. Finally and as an example, Fig.\,\ref{Fig:comparison_with_VanReeth_TAR} compares the TAR obtained by \citet{VanReeth2016} and this work for a particular period-spacing pattern in KIC 7939065. The two results are in agreement and the slight offset of some peaks generates from the coarse parameter space for $\alpha_g$ used in this work and defined in Table \ref{Tab:parameter_space}.

Overall, we conclude from Figs.\,\ref{Fig:comparison_with_VanReeth}, \ref{Fig:comparison_with_VanReeth_and_Ouazzani} and \ref{Fig:comparison_with_VanReeth_TAR} that our method for obtaining the interior rotation frequency and buoyancy travel time works well. Furthermore, the uncertainties we deduced from our bootstrap method are comparable to those in the literature deduced from higher-quality {\it Kepler\/} data. By comparing the median uncertainties obtained from 1-yr TESS and 4-yr {\it Kepler\/} light curves, we conclude that period-spacing patterns from 1-yr TESS light curves can constrain the internal rotation frequency and buoyancy travel time to a precision of 0.03\,d$^{-1}$ and 400\,s, respectively, which is about half as good as the values deduced in the literature and by our re-analyses from 4-yr {\it Kepler\/} light curves. Therefore, the values and uncertainties for $f_{\rm rot}$ and $\Pi_0$ deduced from the 1-yr TESS light curves are robust.

%--------------------------------------
\section{Conclusions}\label{Sec:discussion}
%--------------------------------------

We provided rotation frequencies and buoyancy travel times for a sample of 60 $\gamma$~Dor stars observed in the TESS S-CVZ during 352\,d covered in Cycle\,1 of the mission. These two important observables of stellar interiors can readily be decoded from period-spacing patterns of identified g-mode pulsators. 
While we showed that working with 1-yr light curve comes with its drawbacks, mainly due to missing mode periods in the patterns hampering secure mode identification, the proper information could be deduced for 60 of the 106 $\gamma$~Dor stars in our sample.

We validated our methodology by reproducing {\it Kepler\/} results from \citet{VanReeth2016} and concluded that period-spacing patterns from 1-yr TESS light curves can constrain the internal rotation frequency and buoyancy travel time to a precision of 0.03\,d$^{-1}$ and 400\,s, respectively.

The 60 g-mode pulsators in the TESS S-CVZ made available in this work offer excellent asteroseismic potential to constrain their internal mixing profiles and angular momentum transport, once additional TESS data from the ongoing cycles are added to the light curves. Indeed, the scientific value of the stars in our sample will keep increasing as the TESS mission continues beyond its nominal mission. Besides the asteroseismic dimension, this sample has spectroscopic values for $\sim60\%$ of its stars from spectra available with one or more epochs indicated in Table \ref{Tab:results}. With the third data release of Gaia newly available, this catalogue is a prime candidate for combined asteroseismic, spectroscopic and astrometric modelling to test theoretical and computational tools from generalised TAR and radiative levitation.

%--------------------------------------------------------------------

\begin{acknowledgements}

The research leading to these results has received funding from the  the KU\,Leuven Research Council (grant C16/18/005: PARADISE) and from the BELgian federal Science Policy Office (BELSPO) through PRODEX grants for the Gaia and PLATO space missions. TVR gratefully acknowledges support from the Research Foundation Flanders (FWO) under grant agreement N$^\circ$12ZB620N. We thank the anonymous referee for useful comments which have allowed us to improve the text.

\end{acknowledgements}

%-----------------------------------------------------------------

  \bibliographystyle{aa} 
  \bibliography{ref}

%----------------------------------------------------------------------------
 \begin{appendix}

\clearpage

\onecolumn

\section{Extended tables}

\renewcommand{\arraystretch}{1.4}
\tabcolsep=4pt

\begin{longtable}{rrrrcccrrrrrcc}
\caption{\label{Tab:results}TIC stars in our catalogue and their period-spacing pattern parameters, derived for each pattern that was successfully modelled.} \\
\hline \hline
TIC       & $m_{\rm\small TESS}$  &  $\langle P \rangle$ & $\langle \Delta P \rangle$ & Patterns & $f_{\rm rot}$    & $\Pi_0$ & $k$  & $m$ & $n_{\rm\small long}$ & Periods & $n$-span & Hybrid & Spec \\
          & (mag) & (d)   & (ks)                        &          & ($\rm d^{-1}$) & (s)          &            &    &    &         &        &       &  \\
\hline
\endfirsthead
\caption{continued.}\\
\hline\hline
TIC       & $m_{\rm\small TESS}$  &  $\langle P \rangle$ & $\langle \Delta P \rangle$ & Patterns & $f_{\rm rot}$    & $\Pi_0$ & $k$  & $m$ & $n_{\rm\small long}$ & Periods & $n$-span & Hybrid & Spec \\
          & (mag) & (d)   & (ks)                        &          & ($\rm d^{-1}$) & (s)          &            &    &    &         &        &       &  \\
\hline
\endhead
\hline
\endfoot
38515566  & 8.85  & 1.551 & 2.91                        & 1        & 0.197$^{+0.002}_{-0.025}$             & 3420$^{+50}_{-50}$                      & 0  & -1 & 42 & 12      & 16     & Y  &  1    \\
40335866  & 9.72  & 0.26  & 0.46                        & 1        & 2.760$^{+0.005}_{-0.007}$             & 5820$^{+250}_{-500}$                      & 0  & 1  & 9 & 10      & 14     & Y  &  1  \\
41483281  & 10.25 & 0.617 & 2.76                        & 1        & 0.33$^{+0.02}_{-0.02}$             & 5220$^{+250}_{-250}$                      & 1  & 0  & 9 & 9       & 15     & Y  &  1  \\
55453219  & 8.84  & 0.396 & 0.53                        & 1        & 1.746$^{+0.005}_{-0.012}$             & 5620$^{+100}_{-100}$                      & 0  & 1  & 16 & 5       & 12     &    &      \\
140511383 & 11.92 & 0.436 & 0.45                        & 1        & 1.578$^{+0.015}_{-0.005}$             & 4660$^{+100}_{-100}$                      & 0  & 1  & 19 & 9       & 20     &    &  1   \\
141479660 & 11.23 & 0.465 & 1.33                        & 1        & 0.86$^{+0.03}_{-0.03}$             & 3980$^{+100}_{-200}$                      & 0  & 1  & 12 & 12      & 18     & Y  &  3   \\
149540525 & 8.39 & 0.625 & 0.94                         & 1        & 0.91$^{+0.02}_{-0.02}$             & 5260$^{+200}_{-200}$                      & 0 & 1 & 18 & 13       & 21     & Y     &  3    \\
149573437 & 9.48  & 0.282 & 0.50                        & 1        & 2.31$^{+0.01}_{-0.01}$             & 4280$^{+100}_{-50}$                      & 0  & 1  & 14 & 6       & 8      & Y  &  1   \\
149630117 & 9.05  & 0.392 & 1.47                        & 1        & 1.88$^{+0.06}_{-0.06}$             & 4020$^{+150}_{-200}$                      & 1  & 0  & 12 & 5       & 8      & Y  &      \\
149993830 & 8.91  & 0.551 & 0.48                        & 1        & 2.40$^{+0.05}_{-0.05}$             & 2400$^{+100}_{-200}$                      & -2 & -1 & 12 & 6       & 8      &    &      \\
150165657 & 8.66  & 0.411 & 1.02                        & 1        & 0.00$^{+0.01}_{-0.00}$             & 3520$^{+100}_{-100}$                      & 0  & 2  & 29 & 11      & 13     & Y  &      \\
150318672 & 10.16 & 0.296 & 0.36                        & 1        & 1.05$^{+0.02}_{-0.02}$             & 5200$^{+500}_{-500}$                      & 0  & 2  & 25 & 6       & 8      &    &      \\
150324086 & 10.13 & 0.516 & 0.66                        & 1        & 1.107$^{+0.015}_{-0.015}$             & 3660$^{+150}_{-200}$                      & 0  & 1  & 26 & 8       & 12     &    &      \\
150392753 & 8.52  & 0.304 & 0.17                        & 1        & 2.72$^{+0.02}_{-0.01}$             & 5980$^{+100}_{-500}$                      & 0  & 1  & 17 & 11      & 22     &    &  1   \\
150440102 & 11.72 & 0.933 & 1.38                        & 1        & 2.99$^{+0.06}_{-0.03}$             & 2220$^{+100}_{-100}$                      & -1 & -1 & 38 & 6       & 11     &    &  1   \\
150440362 & 8.02  & 0.870  & 1.31                       & 1        & 1.02$^{+0.06}_{-0.06}$             & 4540$^{+600}_{-500}$                      & 1  & 0  & 29 & 5       & 9      &    &      \\
176874440 & 11.10  & 1.159 & 2.50                       & 1        & 0.74$^{+0.06}_{-0.06}$             & 3120$^{+100}_{-200}$                      & 0  & -1 & 36 & 6       & 6      &    &  1  \\
176980185 & 10.86 & 0.628 & 1.00                        & 1        & 0.93$^{+0.01}_{-0.01}$             & 5960$^{+300}_{-200}$                      & 0  & 1  & 20 & 7       & 9      &    &  1  \\
177082055 & 8.28  & 1.149 & 1.78                        & 1        & 1.00$^{+0.06}_{-0.06}$             & 2200$^{+50}_{-50}$                      & 0  & -1 & 47 & 7       & 16     & Y  &      \\
177115672 & 11.70  & 0.806 & 0.50                       & 1        & 0.803$^{+0.015}_{-0.015}$             & 4080$^{+250}_{-250}$                      & 0  & 1  & 45 & 8       & 16     &    &  1   \\
177164485 & 10.41 & 0.707 & 0.94                        & 1        & 0.00$^{+0.01}_{-0.00}$             & 2280$^{+100}_{-100}$                      & 0  & 2  & 59 & 10      & 15     &    &  1  \\
177386428 & 9.19  & 0.269 & 0.32                        & 1        & 2.82$^{+0.06}_{-0.06}$             & 5860$^{+200}_{-200}$                      & 0  & 1  & 15 & 6       & 7      & Y  &      \\
231084221 & 10.26 & 0.482 & 0.97                        & 1        & 1.00$^{+0.06}_{-0.06}$             & 3820$^{+500}_{-500}$                      & 0  & 1  & 22 & 5       & 5      & Y  &  1  \\
257721280 & 9.36  & 0.860  & 0.99                       & 1        & 0.06$^{+0.01}_{-0.01}$             & 2940$^{+200}_{-200}$                      & 0  & 2  & 55 & 10      & 27     &    &      \\
260265631 & 10.21 & 0.666 & 1.56                        & 1        & 1.24$^{+0.06}_{-0.06}$             & 4840$^{+150}_{-150}$                      & 1  & 0  & 18 & 6       & 10     &    &      \\
260502142 & 10.58 & 0.328 & 0.32                        & 1        & 0.86$^{+0.05}_{-0.05}$             & 3520$^{+600}_{-600}$                      & 0  & 2  & 33 & 9       & 17     & Y  &  1   \\
260540780 & 10.30  & 0.382 & 0.32                       & 1        & 1.78$^{+0.06}_{-0.06}$             & 3180$^{+300}_{-300}$                      & 0  & 1  & 30 & 8       & 11     &    &      \\
271639931 & 8.08  & 0.380  & 0.33                       & 1        & 0.69$^{+0.07}_{-0.07}$             & 3020$^{+600}_{-650}$                      & 0  & 2  & 48 & 4       & 6      &    &  2   \\
279055960 & 11.02 & 0.329 & 0.24                        & 1        & 2.35$^{+0.02}_{-0.02}$             & 4700$^{+100}_{-100}$                      & 0  & 1  & 21 & 6       & 15     &    &  1   \\
279360930 & 11.30  & 0.290  & 0.66                      & 1        & 1.944$^{+0.025}_{-0.025}$             & 3600$^{+150}_{-100}$                      & 0  & 1  & 15 & 5       & 7      & Y  &      \\
279510278 & 11.44 & 0.745 & 0.72                        & 1        & 1.18$^{+0.06}_{-0.06}$             & 2400$^{+150}_{-100}$                      & 1  & 0  & 49 & 8       & 9      &    &  1   \\
293345700 & 9.95  & 0.461 & 0.26                        & 1        & 0.51$^{+0.025}_{-0.01}$             & 2020$^{+300}_{-100}$                      & 0  & 2  & 81 & 6       & 8      &    &  1  \\
293974233 & 11.32 & 0.466 & 2.69                        & 1        & 0.06$^{+0.06}_{-0.06}$             & 3920$^{+200}_{-100}$                      & 0  & 1  & 11 & 8       & 9      & Y  &      \\
294092361 & 10.56 & 1.070  & 0.79                       & 1        & 1.209$^{+0.025}_{-0.020}$             & 4280$^{+500}_{-450}$                      & -2 & -1 & 17 & 7       & 7      &    &  1  \\
300033585 & 9.99  & 0.378 & 1.06                        & 1        & 1.30$^{+0.01}_{-0.01}$             & 4360$^{+100}_{-100}$                      & 0  & 1  & 13 & 6       & 8      & Y  &  1  \\
300138080 & 11.64 & 0.529 & 0.70                        & 1        & 0.98$^{+0.02}_{-0.02}$             & 3120$^{+150}_{-100}$                      & 0  & 1  & 25 & 7       & 19     &    &  1  \\
349092320 & 9.15  & 0.751 & 1.07                        & 1        & 0.56$^{+0.06}_{-0.06}$             & 3380$^{+500}_{-550}$                      & 0  & 1  & 29 & 12      & 22     &    &  1   \\
349096085 & 11.22 & 0.438 & 2.00                        & 1        & 0.72$^{+0.03}_{-0.02}$             & 4820$^{+150}_{-200}$                      & 0  & 1  & 10 & 6       & 10     & Y  &  1   \\
349310718 & 10.52 & 0.408 & 0.22                        & 1        & 1.88$^{+0.01}_{-0.01}$             & 4360$^{+200}_{-200}$                      & 0  & 1  & 31 & 8       & 13     &    &  3  \\
349521873 & 8.96  & 0.804 & 1.56                        & 1        & 0.49$^{+0.06}_{-0.06}$             & 4540$^{+700}_{-750}$                      & 0  & 1  & 24 & 7       & 12     &    &      \\
349683884 & 10.07 & 0.394 & 0.83                        & 1        & 1.41$^{+0.02}_{-0.02}$             & 4380$^{+200}_{-200}$                      & 0  & 1  & 16 & 7       & 8      & Y  &  1   \\
349832567 & 10.80  & 0.448 & 0.25                       & 1        & 1.69$^{+0.02}_{-0.02}$             & 4300$^{+200}_{-200}$                      & 0  & 1  & 30 & 12      & 20     &    &      \\
349902873 & 8.62  & 0.287 & 0.33                        & 1        & 0.84$^{+0.06}_{-0.06}$             & 2600$^{+750}_{-700}$                      & 0  & 2  & 34 & 7       & 15     & Y  &      \\
350092538 & 10.12 & 0.596 & 0.26                        & 1        & 0.55$^{+0.01}_{-0.01}$             & 4600$^{+300}_{-300}$                      & 0  & 2  & 63 & 5       & 17     &    &  1  \\
350144657 & 10.81 & 0.513 & 0.27                        & 1        & 1.47$^{+0.01}_{-0.01}$             & 4640$^{+200}_{-100}$                      & 0  & 1  & 30 & 15      & 25     &    &  1  \\
350343297 & 9.00  & 0.538 & 0.78                        & 1        & 0.025$^{+0.030}_{-0.005}$             & 2000$^{+600}_{-200}$                      & 0  & 2  & 48 & 9       & 21     & Y  &  1   \\
350477538 & 11.34 & 0.356 & 1.01                        & 1        & 0.01$^{+0.03}_{-0.01}$             & 2480$^{+300}_{-50}$                      & 0  & 2  & 29 & 4       & 5      &    &  1  \\
350715741 & 10.87 & 0.545 & 0.59                        & 1        & 1.197$^{+0.020}_{-0.015}$             & 4900$^{+200}_{-150}$                      & 0  & 1  & 18 & 7       & 28     & Y  &  1  \\
350840969 & 11.36 & 0.693 & 0.58                        & 1        & 0.89$^{+0.06}_{-0.06}$             & 4120$^{+700}_{-700}$                      & 0  & 1  & 36 & 6       & 13     &    &  1  \\
364325752 & 9.98  & 0.597 & 0.28                        & 1        & 2.95$^{+0.02}_{-0.07}$             & 5940$^{+200}_{-700}$                      & 2  & 0  & 85 & 7       & 15     &    &  1   \\
388131027 & 8.03  & 0.947 & 1.1                         & 1        & 0.55$^{+0.01}_{-0.02}$             & 5020$^{+600}_{-700}$                      & 0  & 1  & 32 & 11      & 13     &    &      \\
391894459 & 9.31  & 0.436 & 1.3                         & 1        & 0.05$^{+0.04}_{-0.05}$             & 3440$^{+300}_{-200}$                      & 0  & 2  & 23 & 7       & 11     & Y  &      \\
407661375 & 10.04 & 0.715 & 0.46                        & 1        & 1.78$^{+0.06}_{-0.06}$             & 2700$^{+200}_{-300}$                      & -2 & -1 & 19 & 5       & 10     & Y  &  1   \\
381950897 & 9.56 & 0.764 & 0.54                         & 2        & 1.614$^{+0.008}_{-0.006}$             & 3960$^{+40}_{-80}$                      & -2 & -1 & 16 & 12       & 13     &    &      \\
381950897 & 9.56 & 0.479 & 0.22                         & 2        & 1.603$^{+0.015}_{-0.015}$             & 4080$^{+500}_{-500}$                      & 0 & 1 & 34 & 8       & 16     &       &    \\
374944608 & 9.90 & 0.439 & 0.39                         & 2        & 1.510$^{+0.015}_{-0.015}$             & 3520$^{+150}_{-150}$                      & 0 & 1 & 25 & 15       & 22     &     &  3     \\
349835272 &  9.76 & 0.572 & 3.23                        & 2        & 0.45$^{+0.06}_{-0.06}$             & 3880$^{+200}_{-300}$                      & 0 & -2 & 17 & 7       & 11     &      &  1    \\
293273274 & 11.12 & 0.623 & 0.68                        & 2        & 0.10$^{+0.03}_{-0.02}$             & 2100$^{+200}_{-100}$                      & 0 & 2 & 57 & 8       & 28     &       &  3     \\
293273274 & 11.12 & 1.009 & 1.06                        & 2        & 0.007$^{+0.005}_{-0.07}$             & 2800$^{+100}_{-100}$                      & 2 & 0 & 77 & 7       & 13     &       &  3     \\
293221812 & 11.20 & 0.346 & 0.73                        & 2        & 0.17$^{+0.06}_{-0.04}$             & 2200$^{+300}_{-200}$                      & 0 & 2 & 30 & 5       & 13     &       &  1     \\
270503717 & 10.79 & 0.500 & 0.53                        & 2        & 1.27$^{+0.02}_{-0.02}$             & 4160$^{+200}_{-200}$                             & 0 & 1 & 24 & 11       & 17     &       &      \\  
391744540 & 8.69 & 0.502 & 0.44                         & 2        & 1.39$^{+0.06}_{-0.06}$             & 4940$^{+500}_{-600}$                             & 0 & 1 & 24 & 6       & 15     &  Y     &        
\end{longtable}
\tablefoot{TIC stands for the TESS Input Catalogue identifier, while the column $m_{\rm TESS}$ is the  magnitude in the wide-band TESS filter. The columns $\langle P \rangle$ and $\langle \Delta P \rangle$ are the mean period and the mean period spacing, respectively, from \citet{Garcia2022}. The column \textit{Patterns} indicates the number of detected period-spacing patterns while $f_{\rm rot}$ and $\Pi_0$ stand for the rotation frequency and buoyancy travel time, respectively. The columns $k$, $m$, $n_{\rm\small long}$ represent the mode identification, where $n_{\rm\small long}$ is the radial order of the mode with the longest  period in the observed pattern. The column \textit{Periods} lists the number of observed mode periods in the pattern. The column $n$-span is the range of overtones in the theoretical pattern fitted to the data. Hybrid p- and g-mode pulsators are marked with an ``Y'' from \citet{Garcia2022}. The column \textit{Spec} shows the number of epochs of spectra available either in the ESO arxiv or in GALAH DR3.}

\renewcommand{\arraystretch}{1.4}
\tabcolsep=4pt
\begin{longtable}{rrrrcccrrrrrcc}
\caption{\label{Tab:results2}TIC stars in our catalogue and their period-spacing pattern parameters. The fit used two patterns simultaneously.} \\
\hline \hline
TIC       & $m_{\rm\small TESS}$  &  $\langle P \rangle$ & $\langle \Delta P \rangle$ & Patterns & $f_{\rm rot}$    & $\Pi_0$ & $k$  & $m$ & $n_{\rm\small long}$ & Periods & $n$-span & Hybrid & Spec \\
          & (mag) & (d)   & (ks)                        &          & ($\rm d^{-1}$) & (s)          &            &    &    &         &        &       &  \\
\hline
\endfirsthead
\caption{continued.}\\
\hline\hline
TIC       & $m_{\rm\small TESS}$  &  $\langle P \rangle$ & $\langle \Delta P \rangle$ & Patterns & $f_{\rm rot}$    & $\Pi_0$ & $k$  & $m$ & $n_{\rm\small long}$ & Periods & $n$-span & Hybrid & Spec \\
          & (mag) & (d)   & (ks)                        &          & ($\rm d^{-1}$) & (s)          &            &    &    &         &        &       &  \\
\hline
\endhead
\hline
\endfoot
381950897 & 9.56 & 0.764 & 0.54                         & 2        & 1.6030$^{+0.0025}_{-0.0025}$             & 4080$^{+30}_{-30}$                      & -2 & -1 & 16 & 12       & 13     &    &      \\
381950897 & 9.56 & 0.764 & 0.54                         & 2        & 1.6030$^{+0.0025}_{-0.0025}$             & 4080$^{+30}_{-30}$                      & 0 & 1 & 38 & 8       & 16     &       &    \\
374944608 & 9.90 & 0.439 & 0.39                         & 2        & 1.49$^{+0.01}_{-0.01}$             & 3400$^{+100}_{-100}$                      & 0 & 1 & 25 & 15       & 22     &     &  3     \\
374944608 & 9.90 & 0.844 & 1.38                         & 2        & 1.49$^{+0.01}_{-0.01}$             & 3400$^{+100}_{-100}$                      & 1 & -1 & 43 & 5       & 6     &      &  3    \\
349835272 &  9.76 & 0.572 & 3.23                        & 2        & 0.32$^{+0.02}_{-0.01}$             & 4700$^{+150}_{-150}$                      & 0 & -2 & 15 & 7       & 11     &      &  1    \\
349835272 &  9.76 & 0.535 & 3.28                        & 2        & 0.32$^{+0.02}_{-0.01}$            & 4700$^{+150}_{-150}$                     & 1 & 0 & 12 & 4       & 4     &      &  1    \\
293273274 & 11.12 & 0.623 & 0.68                        & 2        & 0.21$^{+0.02}_{-0.02}$             & 2800$^{+100}_{-100}$                      & 0 & 2 & 48 & 8       & 28     &       &  3     \\
293273274 & 11.12 & 1.009 & 1.06                        & 2        & 0.21$^{+0.02}_{-0.02}$             & 2800$^{+100}_{-100}$                      & 2 & 0 & 73 & 7       & 13     &       &  3     \\
293221812 & 11.20 & 0.346 & 0.73                        & 2        & 0.43$^{+0.06}_{-0.04}$             & 3340$^{+80}_{-80}$                      & 0 & 2 & 24 & 5       & 13     &       &  1     \\
293221812 & 11.20 & 0.709 & 2.92                        & 2        & 0.43$^{+0.06}_{-0.04}$              & 3340$^{+80}_{-80}$                      & 0 & -1 & 20 & 5       & 8     &       &  1     \\

\end{longtable}
\tablefoot{Columns are the same as in Table \ref{Tab:results}}

\twocolumn

 \end{appendix}

\end{document}